\documentclass[a4paper,fleqn,usenatbib]{mnras}

\usepackage[T1]{fontenc}
\usepackage{ae,aecompl}

\usepackage{graphicx}	% Including figure files
\usepackage{amsmath}	% Advanced maths commands
\usepackage{amssymb}	% Extra maths symbols
\usepackage{hyperref}

\newcommand{\Mpc}{\mbox{ Mpc}}
\def\VEV#1{\left\langle #1\right\rangle} % This is \VEV{x} => <x>
\title[Reionization and Percolation Theory]{Reionization through the lens of percolation theory}

\author[Furlanetto \& Oh]{
Steven R. Furlanetto,$^{1}$\thanks{E-mail: sfurlane@astro.ucla.edu (SRF)}
and S. Peng Oh$^{2}$
\\
$^{1}$Department of Physics \& Astronomy, University of California, Los Angeles, Los Angeles, CA 90095, USA\\
$^{2}$Department of Physics, University of California, Santa Barbara, Santa Barbara, CA 93106, USA\\
}

\date{Accepted XXX. Received YYY; in original form ZZZ}

\pubyear{2015}

\begin{document}
\label{firstpage}
\pagerange{\pageref{firstpage}--\pageref{lastpage}}
\maketitle

\begin{abstract}
The reionization of intergalactic hydrogen has received intense theoretical scrutiny over the past two decades. Here, we approach the process formally as a percolation process and phase transition. Using semi-numeric simulations, we demonstrate that an infinitely-large ionized region abruptly appears at an ionized fraction of $x_i \approx 0.1$ and quickly grows to encompass most of the ionized gas: by $x_i \sim 0.3$, nearly ninety percent of the ionized material is part of this region. Throughout most of reionization, nearly all of the intergalactic medium is divided into just two regions, one ionized and one neutral, and both infinite in extent. We also show that the discrete ionized regions that exist before and near this transition point follow a near-power law distribution in volume, with equal contributions to the total filling factor per logarithmic interval in size up to a sharp cutoff in volume. These qualities are generic to percolation processes, with the detailed behavior a result of long-range correlations in the underlying density field. These insights will be crucial to understanding the distribution of ionized and neutral gas during reionization and provide precise meaning to the intuitive description of reionization as an ``overlap" process.
\end{abstract}


%%%%%%%%%%%%%%%%%%%%%%%%%%%%%%%%%%%%%%%%%%%%%%%%%%

%%%%%%%%%%%%%%%%% BODY OF PAPER %%%%%%%%%%%%%%%%%%

\section{Introduction} \label{intro}

The reionization of neutral hydrogen in the intergalactic medium (IGM) at $z \ga 6$ is one of the landmark events in the early history of structure formation in our Universe (e.g., \citealt{loeb13}). In addition to marking the point at which galaxy formation affected every baryon in the Universe, it also represents the last major phase transition in the history of the Universe and has profound effects on the fuel supply for future generations of galaxies. As such, reionization has been a prime focus of diverse observational efforts including studies of the cosmic microwave background \citep{hinshaw13, planck15-param}, galaxy emission line studies (e.g., \citealt{kashikawa06, fontana10, schenker12, treu12}), and the nascent field of low-frequency radio observations of the redshifted 21-cm line \citep{bowman10, paciga11, parsons14, dillon15, ali15}. As it presents a crucial constraint on galaxy formation, many studies of galaxies at $z \ga 6$ also focus their interpretation on the process of reionization (e.g., \citealt{robertson15, bouwens15-reion}).

One of the most fundamental aspects of the reionization process is the intermingled structure of the ionized and neutral material. \citet{furl04-bub} developed an analytic model to describe the growth of ionized bubbles from clustered sources during reionization. Using excursion set theory and a simple ``photon-counting" criterion to determine whether a region is ionized, their model predicted the mass (or volume) distribution of ionized regions. The  physical principles of that model have since been applied to a number of ``semi-numeric" algorithms to generate realizations of the ionization field from linear or quasilinear theory, including the public codes DexM \citep{mesinger07} and 21cmFAST \citep{mesinger11} amongst others (see \citealt{zahn11}).  Meanwhile, fully numerical simulations have increased in both size and sophistication over the past decade, incorporating radiative transfer and hydrodynamics to study the reionization problem over a wide range of scales (e.g., \citealt{so14, gnedin14, iliev14, trac15}). In particular, the excursion set formalism is generally considered to provide a reasonably accurate picture of the ionized bubbles throughout reionization, and its qualitative predictions have guided much of our intuition about the reionization process.

To this point, the structure of ionized regions has typically been quantified only crudely, through the power spectrum of the ionization field (or the closely-related 21-cm intensity), largely because that is both easy to calculate and relatively easy to observe with low-frequency radio measurements. However, the power spectrum is only a crude representation of this structure, and -- largely thanks to the limitations of current observations -- more sophisticated measures have received little attention. However, the next generation of radio telescopes hopes to begin imaging the ionization field, at least on large scales. It is therefore imperative to understand its structure, and its relationship to the sources driving reionization, on a fundamental level. 

In this paper, we will demonstrate that the naive view of discrete ionized bubbles growing independently of each other throughout reionization, and in particular the size distributions predicted by \citet{furl04-bub}, are qualitatively wrong. Reionization is a \emph{percolation process} -- a phrase that has often been applied to reionization, but without a rigorous understanding of its profound implications for how reionization proceeds.  Here we will examine reionization using the tools of percolation theory and phase transitions in order to illuminate the growth of ionized bubbles throughout the epoch (for a readable introduction to the topic, see \citealt{stauffer94}). We will show that the  structures differ dramatically from a ``discrete bubble" scenario. In the process, we will shed light on the results of several simulations that discovered the complexity of the underlying structures \citep{iliev06-sim, gleser06, lee08, friedrich11, chardin12}.

Percolation theory has been used in astrophysics before. In the 1980s, it provided a model to understand the structure of spiral galaxies in the context of self-propagating star formation, in which the resulting structures showed a transition from deterministic to stochastic behavior (e.g., \citealt{freedman83, freedman84, schulman86,seiden90}). It has also been applied to the study of the large-scale structure in the Universe as a way to characterize voids, walls, and filamentary structure, an analogy explored early in the history of gravitational instability theory \citep{shandarin83} and more recently in the context of modern structure formation theory (e.g., \citealt{sahni97, shandarin06, shandarin10, falck15}). In that case, it is useful in describing the distribution of low-density regions, but the filling factor of those regions (the quantity of most interest in percolation processes) is only indirectly observable through its implications for the galaxy distribution. However, we shall see that many of the results of these studies are relevant to the percolation of ionized structures during reionization. 

This paper is organized as follows. In \S \ref{perc-random} we review some of the fundamental insights of percolation theory. In \S \ref{21cmfast} we describe the semi-numeric code used in our examples. \S \ref{perc-reion} demonstrates that cosmological reionization is a percolation process and illustrates the emerging structures. \S \ref{size-distbn} applies percolation theory to the size distribution of ionized regions, while \S \ref{scaling} reviews the behavior near the percolation transition. Finally, we comment on the relation of this work to other efforts to understand reionization in \S \ref{discussion} and conclude in \S \ref{conc}.

The numerical calculations here assume $\Omega_m = 0.28$, $\Omega_\Lambda = 0.72$, $h=0.7$, and $\sigma_8=0.82$, but our conclusions are very general and do not depend on these specific choices. All distances quoted herein are in comoving units.

\section{Percolation and Phase Transitions} \label{perc-random}

Percolation is the simplest example of a phase transition, and even the most basic such problems have a rich mathematical structure. In this section we will illustrate some of this structure with reference to a simple example.  More comprehensive treatments can be found in the mathematical and condensed matter physics literature (e.g., \citealt{essam80, isichenko92, stauffer94, saberi15}).

\subsection{The Hoshen-Kopelman Algorithm} \label{hk}

The mathematical treatment of percolation requires a rigorous and straightforward definition of a cluster -- in our case, an ionized bubble. To this point in the astronomical literature there have been few attempts to characterize ionized bubbles as \emph{discrete} objects, though \citet{lin15} have recently shown that the watershed algorithm provides a physically-relevant characterization.  Numerical simulations of galaxy formation are of course very interested in the analogous problem of identifying discrete galaxies (and percolation is relevant in this context; e.g., \citealt{more11}), but ionized bubbles present a much more straightforward problem, as they have well-defined boundaries (at least for sources with soft ionizing spectra, so that the H~II regions have sharp edges).

We therefore use the Hoshen-Kopelman algorithm \citep{hoshen76} to partition the ionized medium into discrete ionized bubbles. This algorithm identifies lattice cells that neighbor each other into separate ionized regions without placing any constraints on the resulting geometrical structures. In our case, it operates on a cubical box of $N^3$ cells. By stepping through the lattice, it assigns each ionized cell an integer label, checking each time whether it is a new label, whether the label should be the same as a single neighbor,\footnote{Note that we will consider ionized cells to be part of the same cluster if and only if the two cubes share a face. Thus regions that border each other on a single edge or corner are considered separate. This is known as body-centered cubic site percolation. Most conclusions of percolation theory are independent of the lattice structure.}  or whether it links up two previously separate bubbles. The relabeling is performed on an array of bubble labels, rather than on the lattice itself. Thus only two passes through the lattice are required in order to complete the partitioning, and the Hoshen-Kopelman algorithm is very efficient.\footnote{We use a modified three-dimensional version of the implementation of T. Fricke, publicly available online at https://www.ocf.berkeley.edu/$\sim$fricke/projects/hoshenkopelman/ hoshenkopelman.html. }

Finally, we note that this algorithm is a limiting case of the friends-of-friends algorithms common in cosmological simulations for identifying discrete, gravitationally bound objects, where the ``linking length" is equal to the (fixed) lattice spacing \citep{more11}.

\subsection{The Cubic Lattice} \label{cubic-lattice}

In this subsection we will consider a simple example of a cubic lattice in order to illustrate some of the key features of percolation.  We begin with a cubical volume, divided into $400^3$ cells. We randomly determine whether each cell is ``ionized" by generating a random number\footnote{We use the Mersenne Twister algorithm \citep{matsumoto98} to ensure randomness and avoid periodic effects, as implemented by A. Fog (http://www.agner.org/random/).} and comparing it to a prescribed ionized fraction $x_i$.  We then use the Hoshen-Kopelman algorithm to partition the ionized volume into discrete bubbles and examine their properties. We emphasize that this is not a proper model of reionization (unless that process proceeded purely randomly), but the interpretation procedure is identical to that used below, and for simplicity of language we shall use the term ``ionized" to refer to the filled cells.

%FIGURE: Order parameter for cubic lattice
\begin{figure}
	\includegraphics[width=\columnwidth]{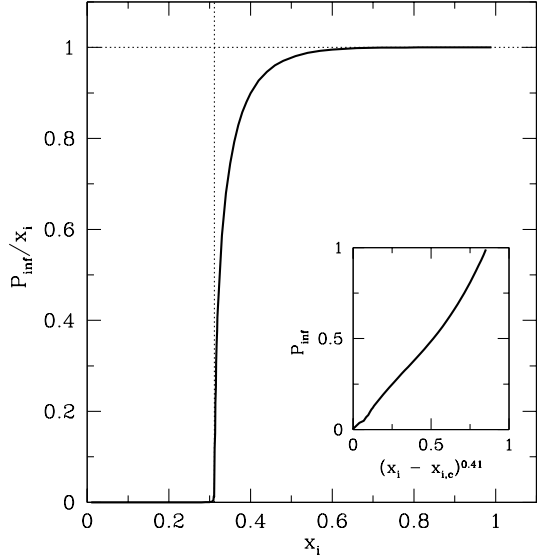}
\vskip -0.9in
    \caption{Order parameter $P_{\rm inf}$ for site percolation on a $400^3$ simple cubic lattice, as a function of the  occupation or ``ionized" fraction $x_i$. We show this in the form $P_{\rm inf}/x_i$, which is the fraction of the ionized volume inside the spanning cluster. The inset shows the universal scaling, $P_{\rm inf} \propto (x_i - x_{i,c})^{0.41}$, for three-dimensional percolation near $x_{i,c}$. For this problem, $x_{i,c} = 0.3116$, marked by the vertical dashed line.}
    \label{fig:order_param_random}
\end{figure}

The most important quality of a percolation process is the phase transition itself, which manifests through the appearance of an infinitely-large ionized region, or \emph{percolating cluster} within the box.\footnote{The infinite nature of this percolating cluster is of course impossible to see in our finite volume. Here and throughout this paper we will identify the percolating cluster as a region that spans the box along at least one of the box's principal axes.}  We will use $P_{\rm inf}$ to denote the fraction of the box's volume in this percolating cluster, or the order parameter; thus $P_{\rm inf}/x_i$ is the fraction of the \emph{ionized} volume within it. Figure~\ref{fig:order_param_random} shows the order parameter for this example. The percolating cluster appears suddenly at a critical threshold $x_{i,c} \approx 0.3116$ and subsequently grows very rapidly, until it includes nearly the entire ionized volume with $x_i \sim 0.5$. \emph{Every} percolation process undergoes such a transition, with an infinite percolating cluster appearing at some well-specified occupation fraction. Moreover, this percolating cluster is (so long as it exists) unique: even in very high dimensions, only one such infinitely large region can exist.  Here we see the first hint that the excursion set treatment of reionization breaks down, for it makes no allowance for such an infinite ionized region.

While the numerical value of $x_{i,c}$ depends on the specifics of the problem at hand, the rapid growth of the percolating cluster itself is a \emph{universal} aspect of the problem.  Near the critical threshold,  
\begin{equation}
P_{\rm inf} \propto |x_i - x_{i,c}|^{\beta},
\label{eq:P_inf}
\end{equation} 
where for a random percolation process $\beta$ is a constant that depends only on the dimensionality of the system (for three dimensions, $\beta=0.41$, as illustrated in the inset to Figure~\ref{fig:order_param_random}). We will see below that the value of $\beta$ is larger in the presence of correlations between cells.

This universal behavior occurs because the system becomes scale-free at the percolation threshold.  This is most apparent if one looks at the cluster (or bubble) size distribution.  We let $dn/dV$ be the number density of these bubbles, per volume in the box (or universe), as a function of each bubble's volume. Percolation theory demands that, at the percolation threshold, $dn/dV \propto V^{-\tau}$, where in three dimensions (without correlations) $\tau \approx -2.18$.  (In detail, this scaling only applies to large $V$, as finite-size effects cause departures for bubbles only a few cells across.) Away from the percolation threshold, this power-law behavior persists at moderately large volumes, but it is cut off at a characteristic scale $V_c$, above which the bubble density declines exponentially fast.  In three dimensions, \citep{stauffer94}
\begin{equation}
n_b(V) \propto 
\begin{cases}
V^{-3/2} \exp^{-C_1 V} & x_i < x_{i,c}, \ \ V \rightarrow \infty \\
V^{1/9} \exp (-C_2 V^{2/3}) & x_i > x_{i,c}, \ \ V \rightarrow \infty,
\end{cases}
\label{eq:perc_size_distbn_scaling}
\end{equation}
for some constants $C_1$ and $C_2$. The gentler scaling above the percolation threshold demonstrates that the abundance is driven by surface effects: more generally, the exponential factor is $\exp (-C_2 V^{1-1/d})$, where $d$ is the dimensionality of the system. 

%FIGURE: Size distribution of clusters for cubic lattice
\begin{figure}
	\includegraphics[width=\columnwidth]{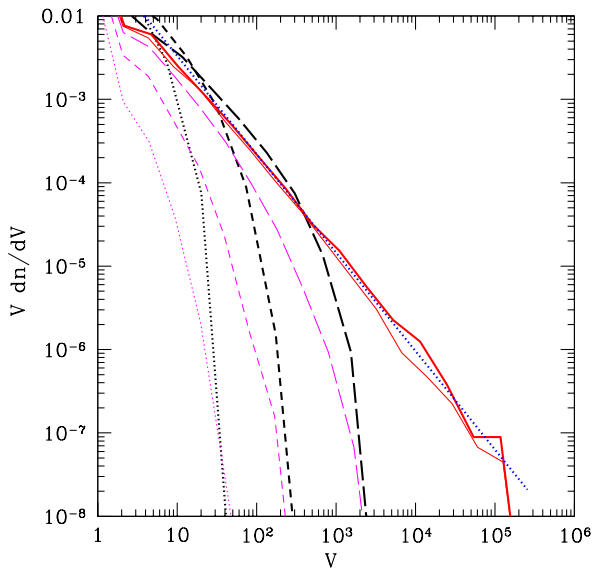}
\vskip -0.9in
    \caption{Size distribution of ``ionized" regions for site percolation on a $400^3$ simple cubic lattice.  The thick and thin solid lines show the distribution just above and just below the percolation threshold, $x_{i,c} = 0.3116$.  The expected scaling at threshold $n_s \propto s^{-2.18}$, is shown by the dotted line.  The thick curves show the distributions below the threshold, while the thin curve show distributions above the threshold (excluding the percolating cluster in the latter case). Note that these all follow the same scaling as the curves at threshold, up to a characteristic scale that declines as $|x_i - x_{i,c}|$ increases.  }
    \label{fig:random_size_distbn}
\end{figure}

Figure~\ref{fig:random_size_distbn} illustrates this scaling in our example box. The solid red lines show the measured distribution just above ($x_i=0.312$) and just below ($x_i=0.311$) the percolation threshold ($x_{i,c}=0.3116$), while the blue dotted line shows the expected $dn/d\ln V \propto V^{-\tau+1}$ scaling.  The other thick lines show the distribution farther below threshold, with $x_i = 0.16,\,0.24,$ and 0.29, from left to right.  The thin curves show the distribution above threshold, with $x_i = 0.46,\,0.38,$ and 0.33, from left to right.  (Recall that the majority of the ionized volume is actually in the unique percolating cluster for $x_i > x_{i,c}$, which is not shown in this figure.) 

Several interesting properties are apparent in Figure~\ref{fig:random_size_distbn} (see \citealt{stauffer94} for a more thorough discussion). First, note how rapidly the cutoff size decreases even slightly away from $x_{i,c}$. This characteristic scale $\xi$ can be defined formally, and its scaling is also universal, with $\xi \propto |x_i - x_{i,c}|^{-\nu}$ for some positive (universal) $\nu$.  If this scale is sufficiently large, the distribution at smaller volumes will follow the power-law slope $\tau$.  Above the cutoff, the decline is gentler above the transition than it is below, in agreement with equation~(\ref{eq:perc_size_distbn_scaling}). 

We will see below that many of these same features appear during reionization, though the correlations intrinsic to that process modify the details.

\section{The 21cmFAST Code} \label{21cmfast}

To generate realizations of reionization, we use the publicly available 21cmFAST semi-numeric code \citep{mesinger11}. This code provides a rapid and reasonably accurate description of the reionization process without performing a full cosmological simulation or any radiative transfer. The ionization structure generated by 21cmFAST differs only in detail from that in full radiative-transfer simulations \citep{zahn11}, though it does not operate on nearly as fine a resolution as some simulations. Our numerical results may therefore differ slightly from simulations, but the generality of percolation theory guarantees that our conclusions are nevertheless robust.

We refer the interested reader to \citet{mesinger11} for details on the code but summarize its main features here. The starting point is a randomly generated linear density field at very high redshift inside a cubical box with periodic boundary conditions, identical to the initial conditions of a cosmological simulation. We generate the density field over 800$^3$ cells with boxes ranging in volume from $100^3$--$400^3$~Mpc$^3$.

This density field is then evolved using linear theory and the Zel'dovich approximation to a later redshift (we will typically use $z=10$). Sources within each cell are then identified using arguments from excursion set theory (specifically, imposing a mass function to match the \citealt{sheth02} model).  The ionization field is also generated using excursion theory arguments, by filtering the sources over larger scales to identify ionized regions, following \citet{furl04-bub}.  We perform this filtering on a coarser scale than that in which the density field is generated, using $100^3$--$400^3$ cells. We use the code's default option of identifying ionized regions on a cell-by-cell basis \citep{zahn11} rather than generating spherical ionized regions as in \citet{mesinger07}. This is slightly less accurate in comparison to numerical simulations, but it is much faster, particularly at the end of reionization (which we will discuss in \S \ref{discussion}).  We note that semi-numeric approaches like 21cmFAST are predicated on the excursion set formalism, which (as we will later show) provides an incomplete picture of the reionization process and structures -- at least in its analytic formulation. However, the fundamental principle of the model, which distributes ionizing photons across many scales and constructs ionized regions by comparing sources and sinks, is sound and agrees well enough with numerical simulations.

We use the default parameter choices of 21cmFAST in almost all of our calculations. In addition to the usual cosmological parameters, these include $R_{\rm max}$, which imposes a maximum size on the sphere of influence for any given cell's ionizing sources and is meant to crudely model absorption within the IGM \citep{furl05-rec}. We set $R_{\rm max}=30$~Mpc, unless otherwise specified. This is likely a very poor representation of IGM absorption \citep{crociani11,sobacchi14,davies15-uvb}; to the extent that improved models reduce large-scale correlations they may have interesting effects on the percolation results and merit further study.

There are two additional subtleties in applying 21cmFAST to percolation problems. First, the relatively low resolution of the simulations (we will typically use cells 1~Mpc across, below which 21cmFAST's treatment of the halo distribution becomes unreliable) means that some fraction of the pixels will only be partially ionized.  At higher resolution, such a cell would contain a mixture of fully-ionized H~II regions and neutral gas.  Such a region could therefore be part of a percolating cluster, so long as those H~II regions connect opposite sides of the cell. These regions should therefore be treated carefully.  Without fine-scale information, we  use a simple criterion to determine whether to treat a cell as fully ionized or fully neutral for the purposes of percolation tests. We order all the partially ionized cells by their ionized fraction and choose a threshold, above which all cells are labeled as fully ionized, so that the box's true total ionized fraction is conserved during the transformation.  This typically corresponds to treating all pixels with $x_i \ga 0.25$ as fully ionized.  We have verified that this choice has no effect on our qualitative and quantitative results.

Second, 21cmFAST -- like nearly all cosmological simulations -- uses periodic boundary conditions in order to self-consistently account for the influence of material ``beyond" the box's edges. This means that structures on opposite sides of the box are not independent of each other, as they were in the example from the previous section. Our numerical experiments therefore have a somewhat smaller ``effective" volume than their nominal value, for the purpose of percolation tests. However, our conclusions are unchanged if we examine sub-volumes of large-scale simulations, so we have not attempted to adjust the boundary conditions.

Because we are concerned with the underlying mathematics of the ionization structure, rather than with reproducing any particular reionization history, we will generally use 21cmFAST at a fixed redshift ($z=10$) and with a fixed set of initial conditions. We then artificially vary the assumed ionizing efficiency of the luminous sources in order to change the ionized fraction (and hence structures).  Each of the curves in our figures are generated in this way: considering a specific realization, at a fixed redshift, over a range of ionized fractions. Fortunately, the ionization field, at a fixed $x_i$, is mostly independent of redshift \citep{furl04-bub, mcquinn07}, so we expect our conclusions to hold even if treated in a time-dependent fashion.  Additionally, we examine the redshift dependence of our results in \S \ref{size-z}.

%FIGURE: Order parameter cosmic variance
\begin{figure}
	\includegraphics[width=\columnwidth]{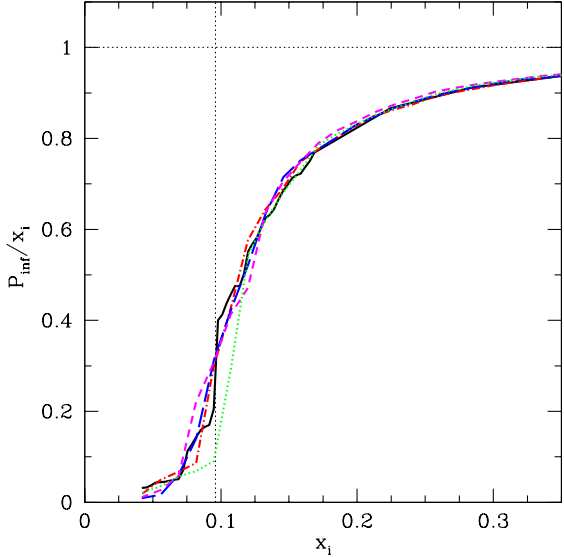}
\vskip -0.9in
    \caption{Order parameter $P_{\rm inf}$ for percolation scaled to the occupation fraction $x_i$, or the fraction of the ``ionized" volume inside the largest cluster. The five curves show results for five independent 200$^3$ boxes spanning $(200 \Mpc)^3$. The dotted vertical curve shows the average percolation threshold, taken to be the point at which the largest cluster first spans the simulation box in at least one dimension. Note that we actually show the fraction of the ionized volume in the \emph{largest} cluster, which is non-zero even below the percolation threshold.}
    \label{fig:order_param_cosvar}
\end{figure}

%FIGURE: Order parameter
\begin{figure}
	\includegraphics[width=\columnwidth]{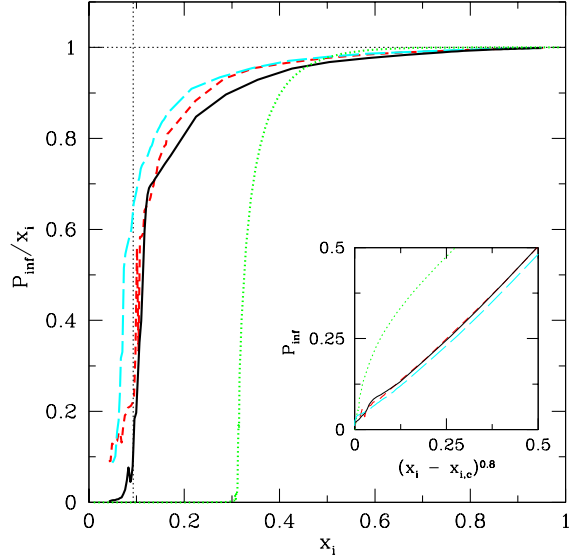}
\vskip -0.9in
    \caption{Order parameter $P_{\rm inf}$ for percolation scaled to the occupation fraction $x_i$, or the fraction of the ``ionized" volume inside the spanning cluster. The solid black curve shows results for a 400$^3$ box spanning $(400 \Mpc)^3$. The red and cyan curves take 100$^3$ and 400$^3$ boxes spanning $(100 \Mpc)^3$. The green curve shows the result for simple cubic site-based interpolation for reference.  The inset illustrates the scaling behavior around threshold for all three reionization scenarios. The dotted vertical curve shows the percolation threshold for the $400^3$ box.}
    \label{fig:order_param}
\end{figure}

\section{Reionization as a Percolation Process} \label{perc-reion}

In the next three sections we will examine how realizations of reionization reflect the expectations of percolation theory. We will consider each of the characteristics identified in \S \ref{perc-random} in turn: the appearance of a percolation threshold, the size distribution of ionized regions, and scaling near the percolation threshold.

%FIGURE: 2D slices
\begin{figure*}
	\includegraphics[width=5.5cm]{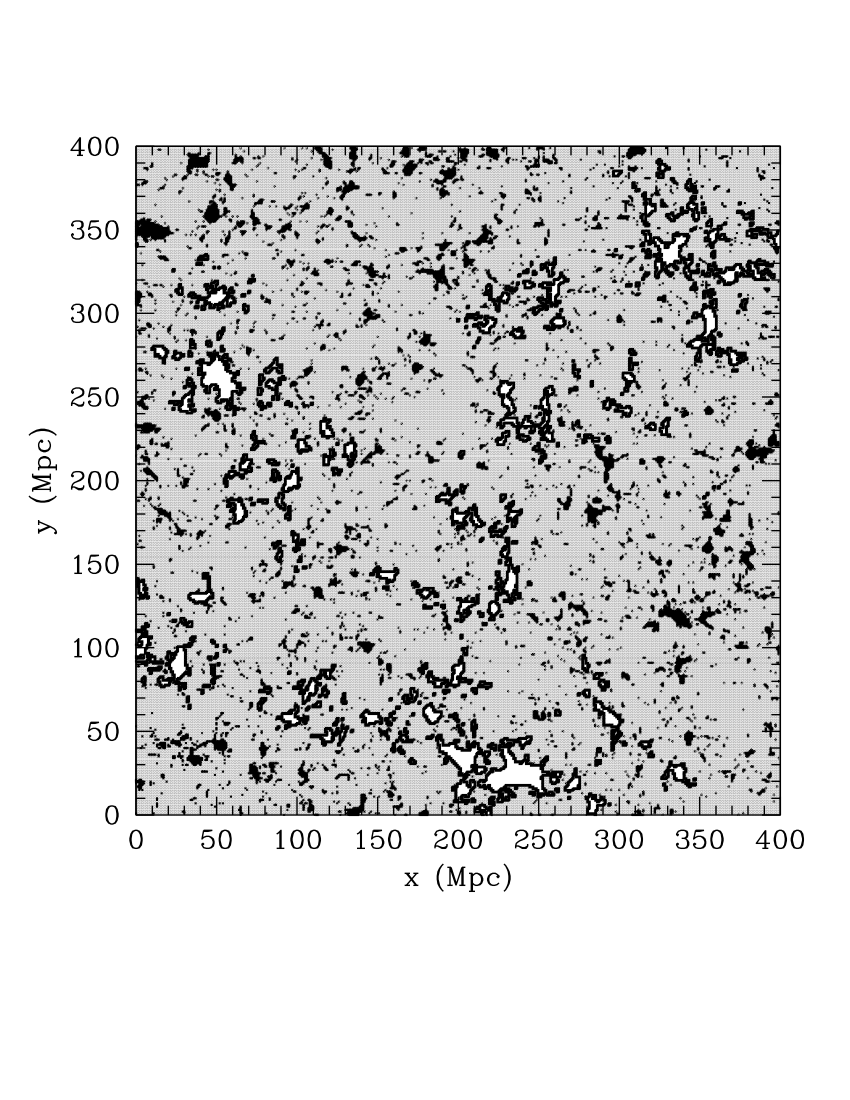} \includegraphics[width=5.5cm]{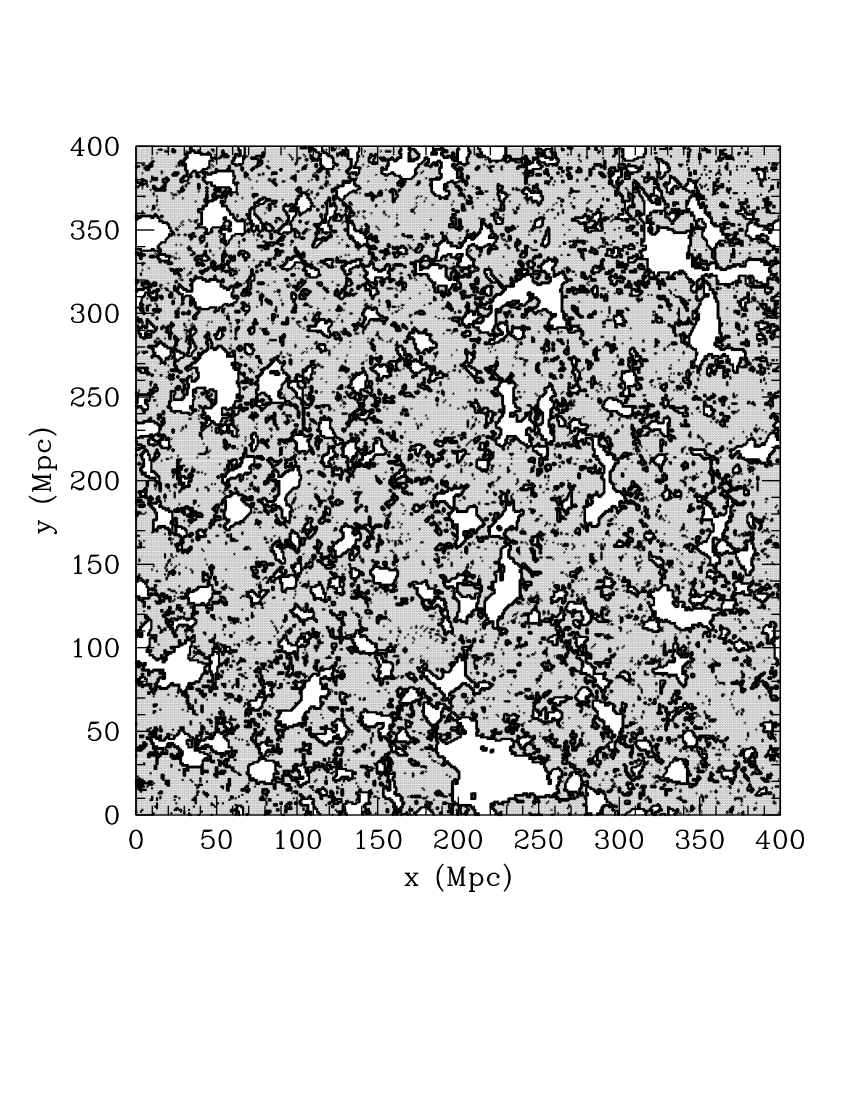} \includegraphics[width=5.5cm]{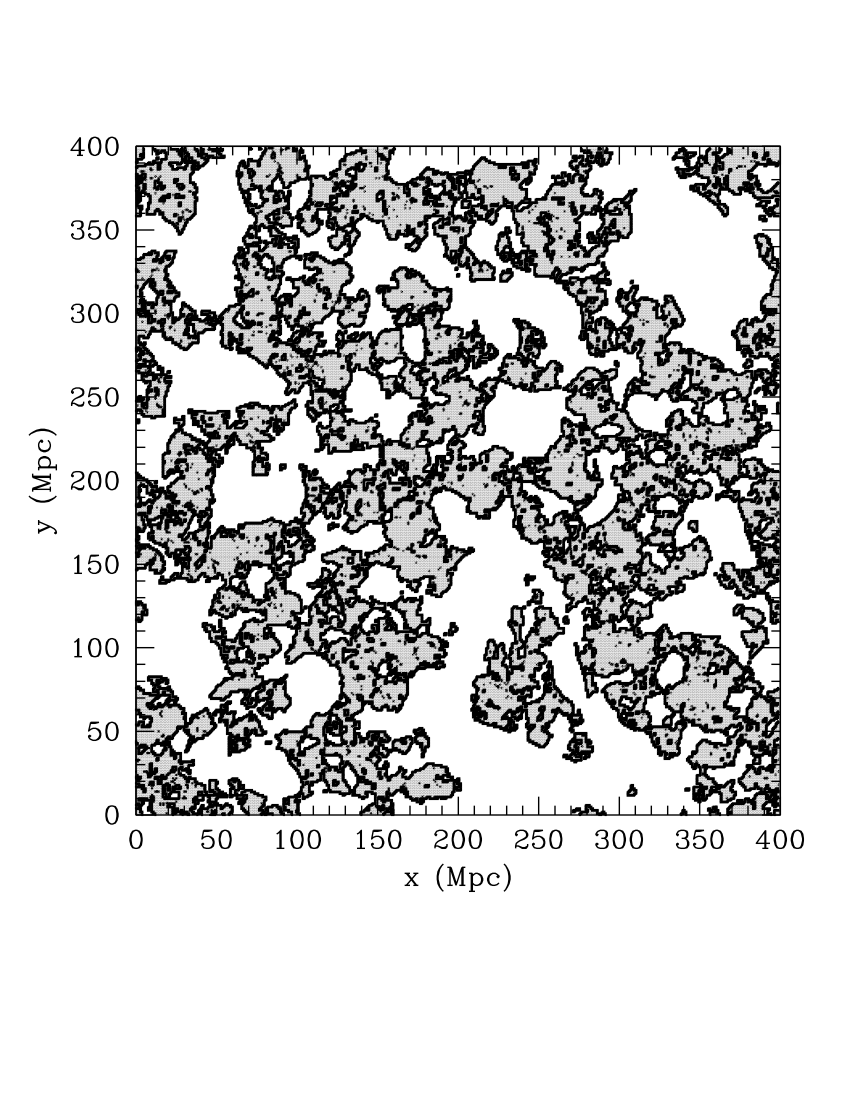}
\vskip -0.5in	
    \caption{Two-dimensional slices through our fiducial 400~Mpc$^3$ box at $x_i\approx0.10,\,0.22$, and $0.50$, from left to right. In each panel, neutral gas is colored  gray, ionized gas that is part of the spanning cluster is colored white with black borders, and ionized gas in separate clusters is solid black. These three slices have been selected to contain the highest fraction of cells in the spanning cluster of all slices perpendicular to the $z$-axis of the box. Note the difficulty of associating connected regions in two-dimensional slices.}
    \label{fig:slices}
\end{figure*}

\subsection{The Order Parameter} \label{order-param}

We begin by examining the appearance of the percolation threshold.  Figure~\ref{fig:order_param_cosvar} shows the order parameter $P_{\rm inf}$ scaled to the total ionized volume.\footnote{Throughout this paper we will use $x_i$ to refer to the \emph{volume-averaged} ionized fraction, as that is the definition most applicable to percolation.}   In our finite box, it is of course impossible to identify whether a region is actually infinite. We therefore have actually plotted the fraction of the ionized volume inside the \emph{largest} ionized region.  We then identify the percolation threshold with the point at which this largest region spans the entire box along at least one of its cardinal directions.\footnote{This criterion is standard in the percolation theory literature, though it is somewhat less meaningful here because of the periodic boundary conditions. However, we have verified that our conclusions are quantitatively unchanged if we repeat our experiments using sub-volumes of larger boxes.}

Figure~\ref{fig:order_param_cosvar} shows results for five independent realizations of the reionization process, each using $200^3$ cells over a $(200 \Mpc)^3$ volume. As expected, $P_{\rm inf}$ rises rapidly and approaches $x_i$ fairly quickly after the percolating bubble appears.  However, there is one key difference from the example in Figure~\ref{fig:order_param_random}, in that the percolation transition occurs far earlier, at $x_i \approx 0.095$ on average. Three of the realizations have the transition occurring at $x_i \approx 0.085$, another at $x_i \approx 0.10$, and one -- the dotted curve -- has it delayed until $x_i \approx 0.12$. This level of variation is not unexpected, given that percolation is an inherently stochastic process. Nevertheless, there is  little variation in $P_{\rm inf}$ between the different realizations, especially past the percolation threshold: the growth of the percolating bubble is remarkably robust.

There are two reasons why $x_{i,c}$ decreases compared to our example cubic lattice. The first is purely geometric: continuum percolation processes (in which the clusters are not limited to exist only on a regular lattice) have somewhat smaller percolation thresholds, as the ionized regions have more freedom in how they can connect to each other. While 21cmFAST is built on a regular lattice, it models such a continuum process.

The more important reason, however, is the clustering of ionized sources: unlike our initial example, the presence of one ionized cell makes the neighboring cells much more likely to be ionized.  Thus in reionization it is much easier to build large ionized structures than it is in a purely random process: this is illustrated by the buildup of a relatively large ionized region even \emph{before} percolation actually occurs (compare to the sharp rise in $P_{\rm inf}$ in Fig.~\ref{fig:order_param_random}). These large, connected ionized bubbles trace the filaments and walls of large-scale structure, allowing percolation to occur much more efficiently than in a random field.  \citet{shandarin06} has found a similarly low percolation threshold with gaussian random fields when studying the percolation of voids in large-scale structure (see \S \ref{lss} below).

\subsection{Convergence Tests} \label{conv}

While Figure~\ref{fig:order_param_cosvar} demonstrates that the percolation transition is robust across different realizations, it does not address the importance of numerical effects like resolution and dynamic range. The effects are not obvious: while smaller boxes will naturally percolate faster in an uncorrelated process, they also have less large-scale power, which helps build the large bubbles. We remedy this in Figure~\ref{fig:order_param}, in which the main panel shows the order parameter across a range of box sizes and resolutions. The solid curve is a $400^3$ grid overlaid on a box of volume $(400 \Mpc)^3$.  It thus has identical resolution to the realizations in Figure~\ref{fig:order_param_cosvar} but eight times the volume.  The vertical dotted curve shows the percolation threshold in this case, at $x_{i,c} \approx 0.10$, very close to the average of the smaller volumes. The short-dashed curve shows a $100^3$ box with volume $(100 \Mpc)^3$; it also has $x_{i,c} \approx 0.09$. The stability of this transition with box size implies that we are not mistaking large bubbles induced by long-range correlations for the percolation threshold.

The long-dashed curve shows our highest resolution box, a $400^3$ box with volume $(100 \Mpc)^3$. This behaves very similarly to the others, except that it has $x_{i,c} \approx 0.07$.  This suggests that we may be slightly overestimating the transition point because our fiducial boxes do not resolve all of the ionized channels generated by the source distribution.  At this level of accuracy, however, a full-scale radiative transfer simulation is necessary for robust results, so we do not pursue this question any further in the present work.  In any case, \emph{the appearance of a large ionized region spanning the simulation volume is a robust feature of the reionization process.}

%FIGURE: Volume rendering
\begin{figure*}
	\includegraphics[width=\columnwidth]{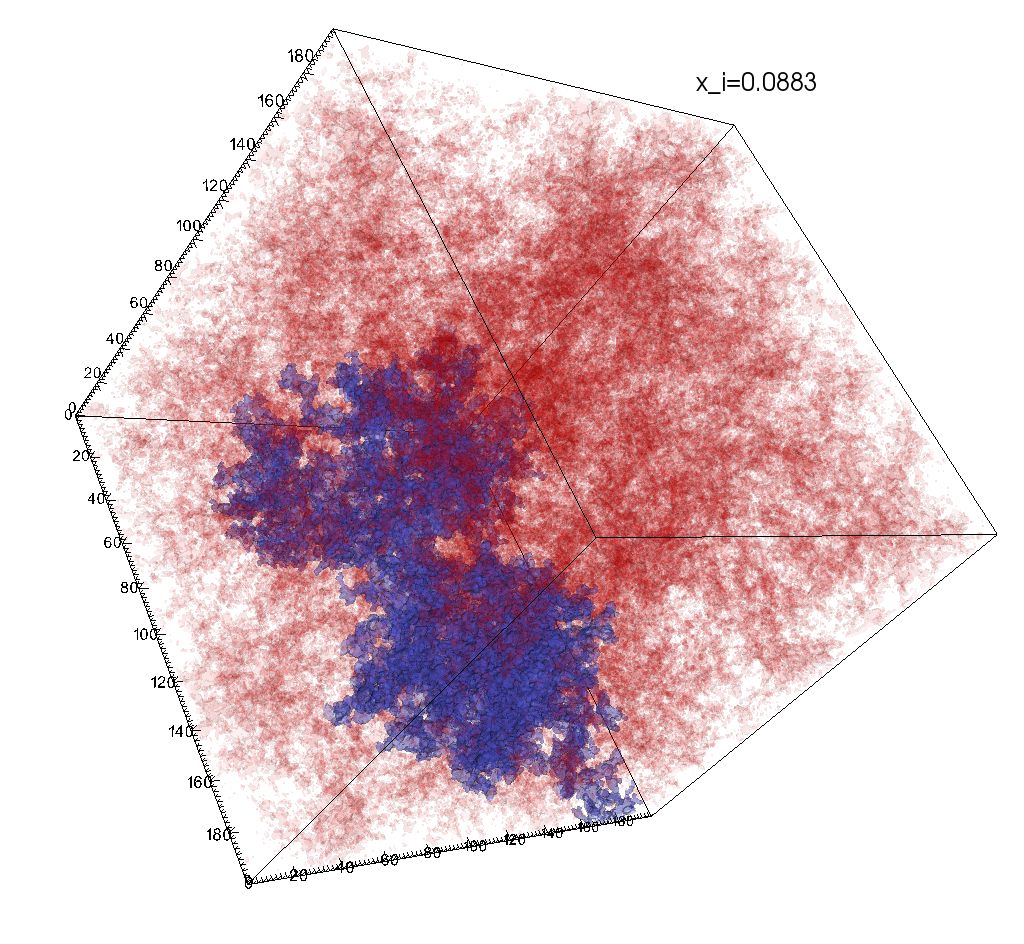} \includegraphics[width=\columnwidth]{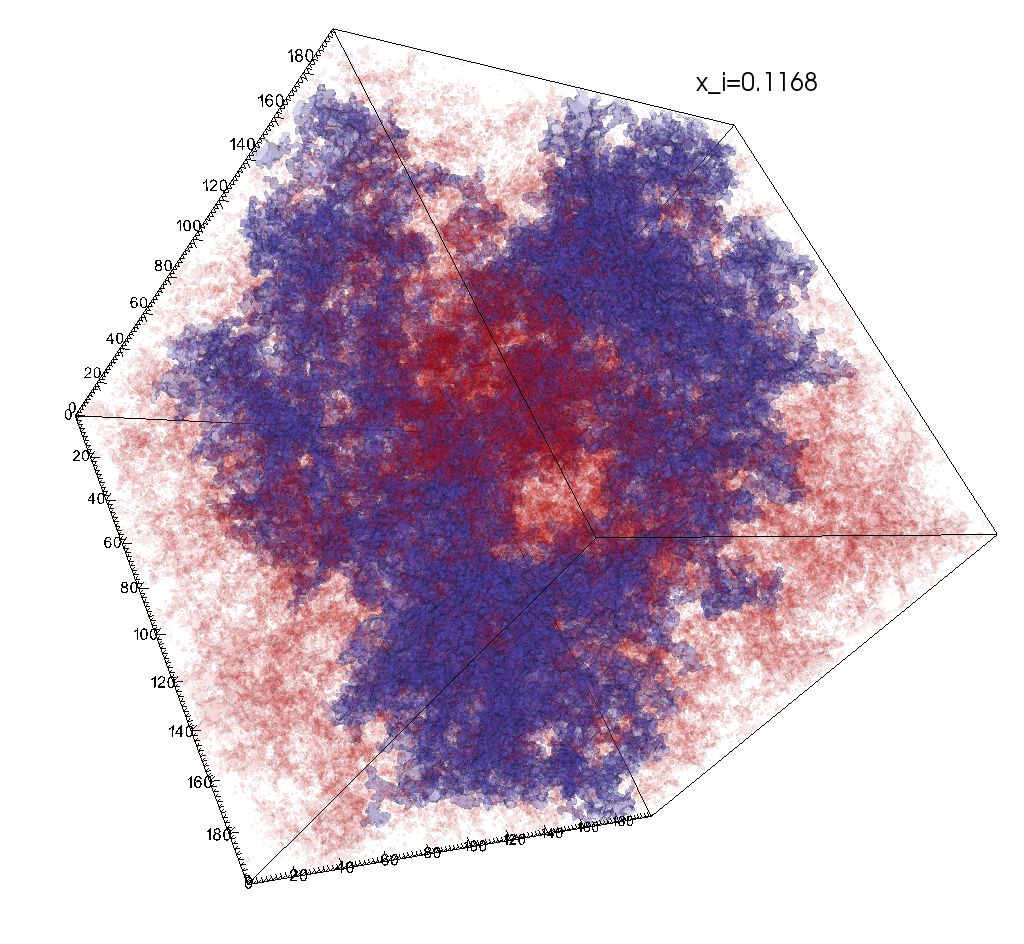} \newline
	\includegraphics[width=\columnwidth]{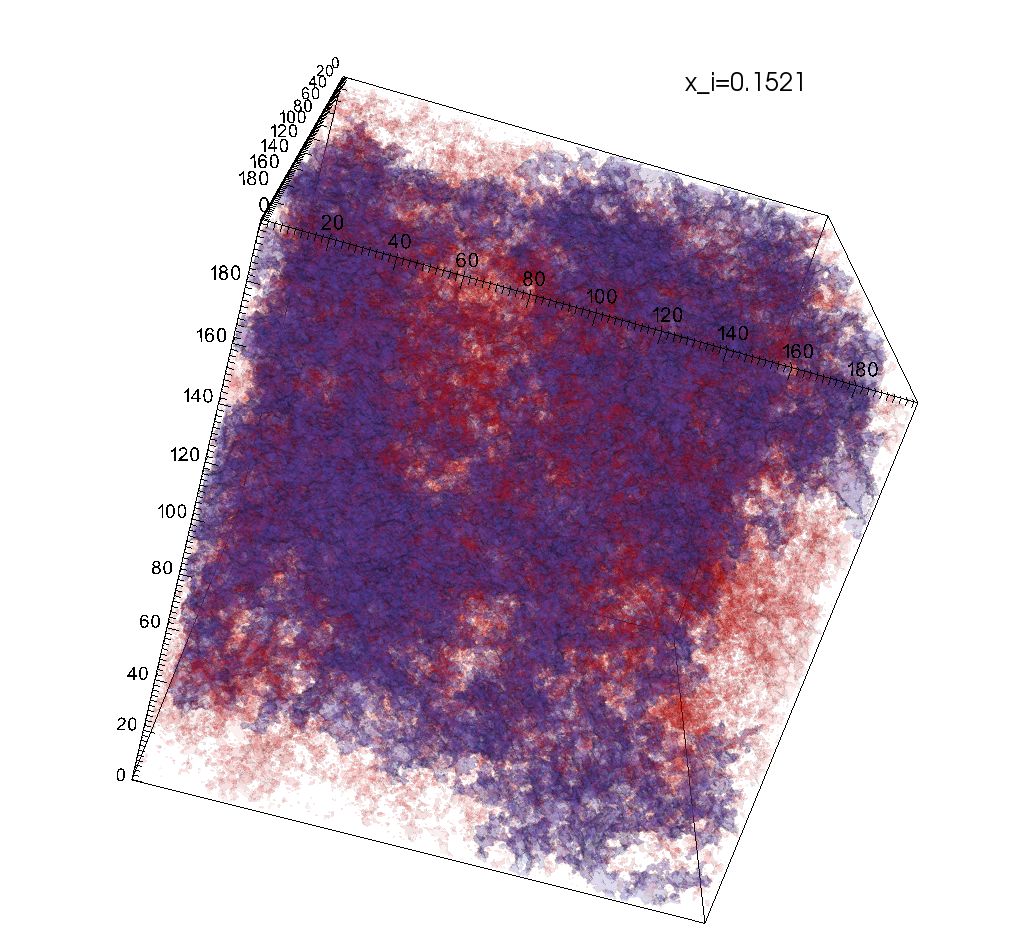} \includegraphics[width=\columnwidth]{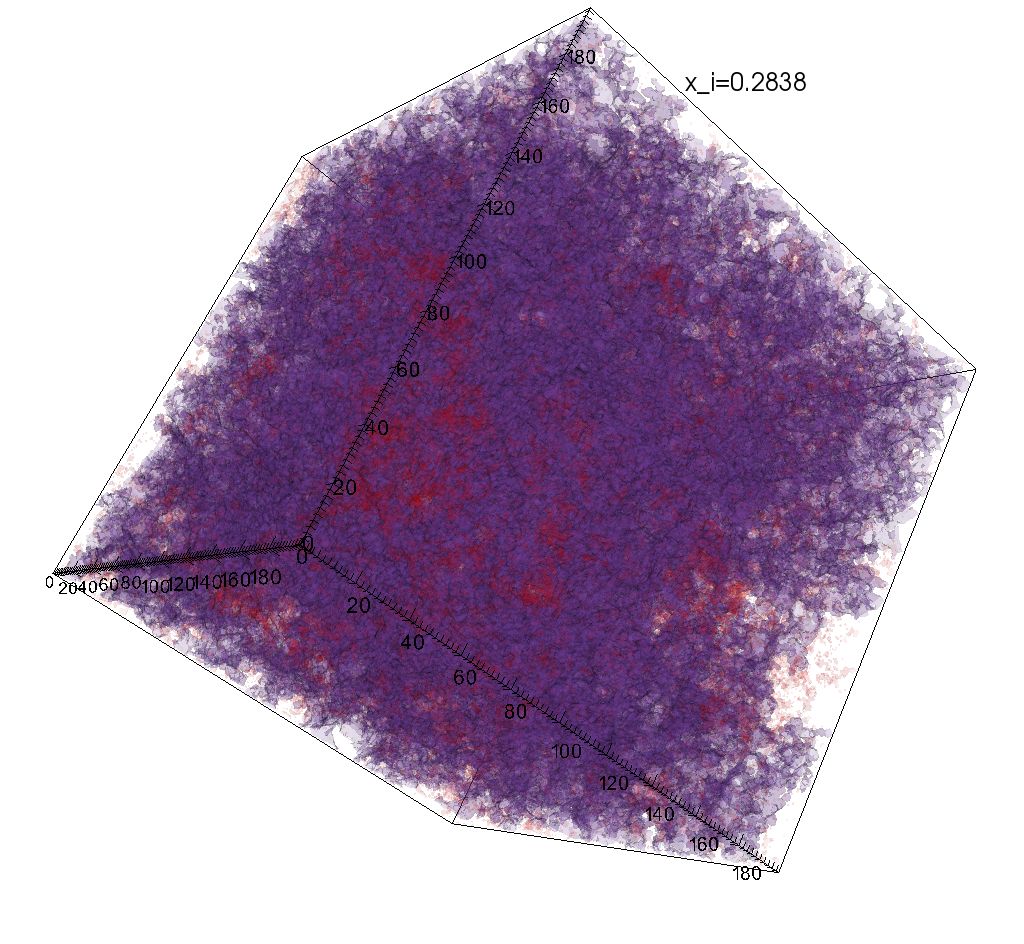} 
\vskip -0.0in	
    \caption{Volume renderings of ionized bubbles in a realization with 200$^3$ cells spanning $(200 \Mpc)^3$. The top panels take $x_i = 0.088$ and 0.12; the former has $x_i \approx x_{i,c}$.  The bottom panels take $x_i = 0.15$ and 0.28, by which point the infinite cluster contains about 90\% of the ionized volume.  Here the blue shows the infinite cluster, while the red show other ionized regions (neutral gas is left transparent). The bottom two panels have been rotated and have had their opacity adjusted for better viewing.}
    \label{fig:rendering}
\end{figure*}

Box size may play a role in another aspect of the problem. In Figure~\ref{fig:order_param_random} (and in all similar problems), the order parameter grows discontinuously at the percolation threshold, making it a first-order phase transition. The curves in Figure~\ref{fig:order_param_cosvar} appear continuous, as do our (100 Mpc)$^3$ simulations in Figure~\ref{fig:order_param}. However, the largest box has a much sharper transition. As we will see in \S \ref{size-distbn}, correlations do induce structure on $\ga 100$~Mpc scales. Resolving the precise nature of the phase transition therefore requires very large boxes. However, such detailed distinctions are very unlikely to be observable, so we do not pursue them further here.

\subsection{Comparison to Previous Work} \label{comp-perc}

The appearance of a percolating cluster has been noted before in numerical simulations of reionization. \citet{iliev06-sim} performed a $(100 h^{-1} \Mpc)^3$ simulation. They found that very large bubbles appeared (with volumes $\ga 10^4 (h^{-1} \Mpc)^3$) once $x_i \ga 0.1$, containing most of the ionized volume, and that this bubble became effectively infinite by about the midway point of reionization. The appearance of these large regions at $x_i \approx 0.1$ is quite striking, though the fact that they did not immediately percolate in this simulation requires some further investigation. One potential discrepancy may come from their different algorithm for identifying discrete bubbles, which was not identical to the Hoshen-Kopelman algorithm (and therefore less forgiving of narrow connections between ionized regions). Alternatively, fine-scale features in the density field may have left narrow neutral channels between the ionized bubbles that persisted much longer than in our crude characterization of IGM absorption. In that case, correlations in the \emph{absorbers} would have a significant effect on the reionization process, but those were not carefully modeled in the \citet{iliev06-sim} simulations.

\citet{chardin12} have also examined the growth of ionized regions during reionization using cosmological simulations. They followed the birth and growth of ionized regions (identified with a friends-of-friends algorithm) throughout reionization. They also found that a single region appeared early in the process (at $x_i \la 0.15$) that quickly dominated the ionized volume. The number of discrete ionized regions peaked at about this same point, with the percolating cluster growing primarily by merging rapidly with these ionized bubbles. Interestingly, they find that at any given moment the bubbles merging with the percolating cluster are roughly characteristic of the overall population of H~II regions, and \citet{chardin14} showed that the time at which any given H~II region is absorbed into the percolating cluster is roughly independent of its source halos' masses. Finally, \citet{chardin12} showed that the precise growth trajectory of the percolating cluster depends on the source model.

We also note that \citet{wang15} have recently compared the topology of the ionized regions between fully-numerical simulations and 21cmFAST, finding differences in certain regimes. It is not clear to what extent these differences were a result of the differing source models in the two cases, as opposed to the algorithms, but it suggests that the details of the percolation process should be explored with more rigorous calculations in the future.

\subsection{Visualizing Percolation} \label{vis}

%FIGURE: Zoomed structure
\begin{figure}
	\includegraphics[width=\columnwidth]{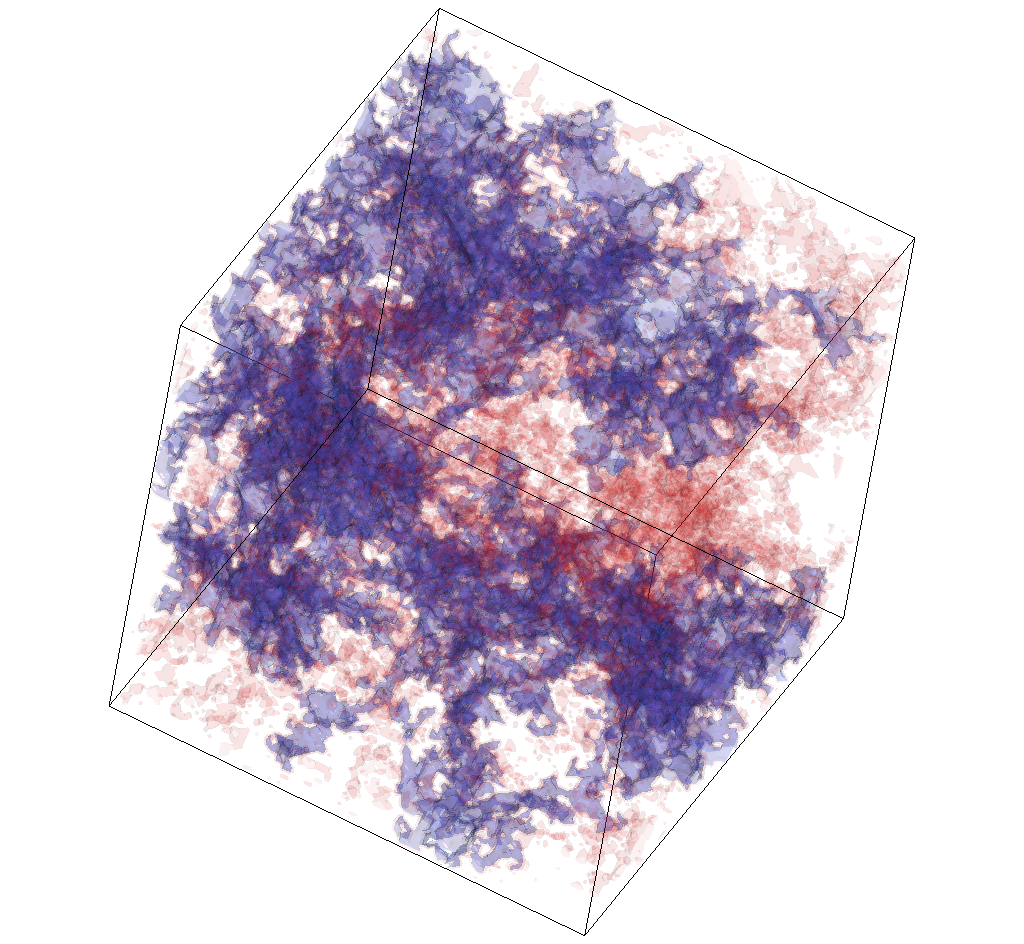}
\vskip -0in
    \caption{Volume rendering of ionized bubbles in a realization with 100$^3$ cells spanning $(100 \Mpc)^3$. This plot has $x_i = 0.12$. The blue shows the infinite cluster, while the red show other ionized regions (neutral gas is left transparent). Note the complex structure of the infinite cluster, which consists of irregular ``blobs" connected by narrow channels. The axes have been omitted so as not to obscure details of the structure.}
    \label{fig:zoom_structure}
\end{figure}

The previous two sections have established that, at least in 21cmFAST simulations, an ionized bubble develops and spans the entire simulation volume very early in the reionization process -- at  $x_i \sim 0.1$.  This has been previously noted by \citet{iliev06-sim} and \citet{chardin12}.  Figure~\ref{fig:slices} illustrates why. These three panels show slices through our $(400 \Mpc)^3$ realization at $x_i\approx0.10,\,0.22$, and $0.50$. The gray is neutral material, the white (with black borders) is ionized material inside the spanning cluster, and the solid black is other ionized material. In each panel, we have chosen the slice, oriented perpendicular to one of the box's axes, that contains the \emph{most} material inside the infinite bubble. Even so, it is impossible to tell from the slice alone that these regions are connected, at least until $x_i \sim 0.5$.  Connections in the orthogonal direction are obviously crucial to identifying reionization as a percolation process.

Figure~\ref{fig:rendering} illustrates this with a series of volume renderings of the ionized regions in a realization with 200$^3$ cells spanning $(200 \Mpc)^3$ (this corresponds to the solid curve in Fig.~\ref{fig:order_param_cosvar}).  The renderings have been prepared with the publicly available code VisIt \citep{visit}.\footnote{See https://wci.llnl.gov/simulation/computer-codes/visit/.}  Here the infinite bubble is outlined in blue, while other ionized regions are shown in red. (The neutral material is left transparent.)  The top left panel shows the structure very close to the percolation threshold, at $x_i \approx 0.088$ in this realization.  The ``infinite" cluster stretches across the box, but it is nevertheless confined to a well-defined subset of the box. Thus it is possible to choose two-dimensional slices that contain \emph{no} elements of the percolating cluster, unlike the left panel of Figure~\ref{fig:slices}.  However, very quickly after this point (in the upper right and especially lower left panels), the percolating cluster connects to other large ionized regions and begins to fill the box.  By $x_i \approx 0.15$, it is impossible to choose a two-dimensional subset whose ionized regions are not dominated by the percolating cluster.  By $x_i \approx 0.25$, the remaining distinct ionized regions are almost all small and isolated from each other -- remnants waiting to be swallowed up by the percolating cluster.

%FIGURE: Shape parameters
\begin{figure}
	\includegraphics[width=\columnwidth]{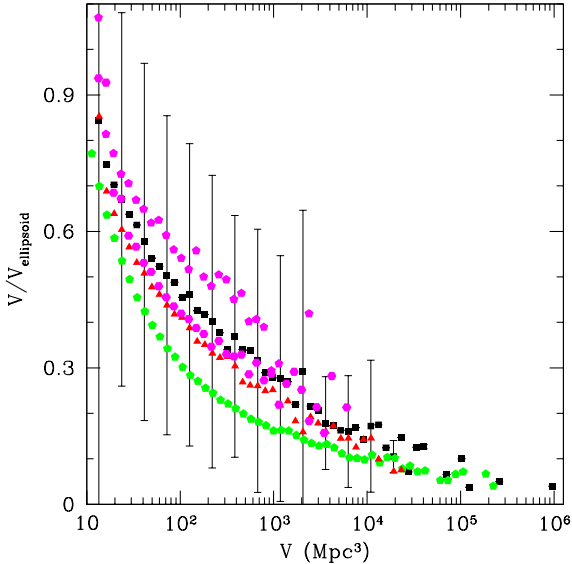}
\vskip -0.9in
    \caption{Inverse porosity of ionized regions in a $400^3$ box spanning (400~Mpc)$^3$.  The  squares show the average porosity as a function of bubble volume near the percolation threshold $x_{i,c} \approx 0.1$. The ``error" bars show the 2$\sigma$ dispersion around the average in selected bins. The triangles and hexagons show the averages for $x_i \approx 0.06$ and $x_i \approx 0.14$, respectively.  The pentagons above the other points show the average values well above threshold, at $x_i \approx 0.35$.  The green pentagons below the other points show the values for random percolation on a similar lattice.  Note that the estimates at $V < 100$~Mpc$^3$ are significantly affected by discreteness in the lattice, and we exclude the largest (percolating) bubble when $x_i > x_{i,c}$.}
    \label{fig:porosity}
\end{figure}

This cluster itself also has a very complex structure, as illustrated by Figure~\ref{fig:zoom_structure}, which shows the ionization field at $x_i \approx 0.12$ in a smaller realization so that the small-scale details are more evident. Note how the cluster twists and wraps upon itself; at higher ionized fractions, in can develop large holes inside of which other large ionized regions can appear. Moreover, the cluster is built from a ``skeleton" of large, irregular regions with many narrow branches leading off rom it, which can themselves branch. This skeleton traces the distribution of luminous sources, or the cosmic web of sheets and filaments.  In fact the percolating cluster (as well as all large clusters) has fractal structure, with features on all scales.  This detailed structure will be important for characterizing large ionized regions in real data with finite resolution, such as 21-cm surveys.

Despite the complexity of the percolating cluster, \citet{lin15} have demonstrated that there are well-defined characteristic scales within it, at least so far as it is defined in a manner other than the Hoshen-Kopelman algorithm (or the related friends-of-friends technique). Measurements based either on the mean length of rays cast from ionized regions to neutral gas or on the watershed algorithm have clear peaks in the distribution of bubble ``sizes," though they differ from the naive expectations of \citet{furl04-bub}.  These characteristic scales capture the physics of source formation, IGM absorption, and the percolation process and are therefore crucial to understand in the future.

%FIGURE: Size distribution
\begin{figure*}
	\includegraphics[width=\columnwidth]{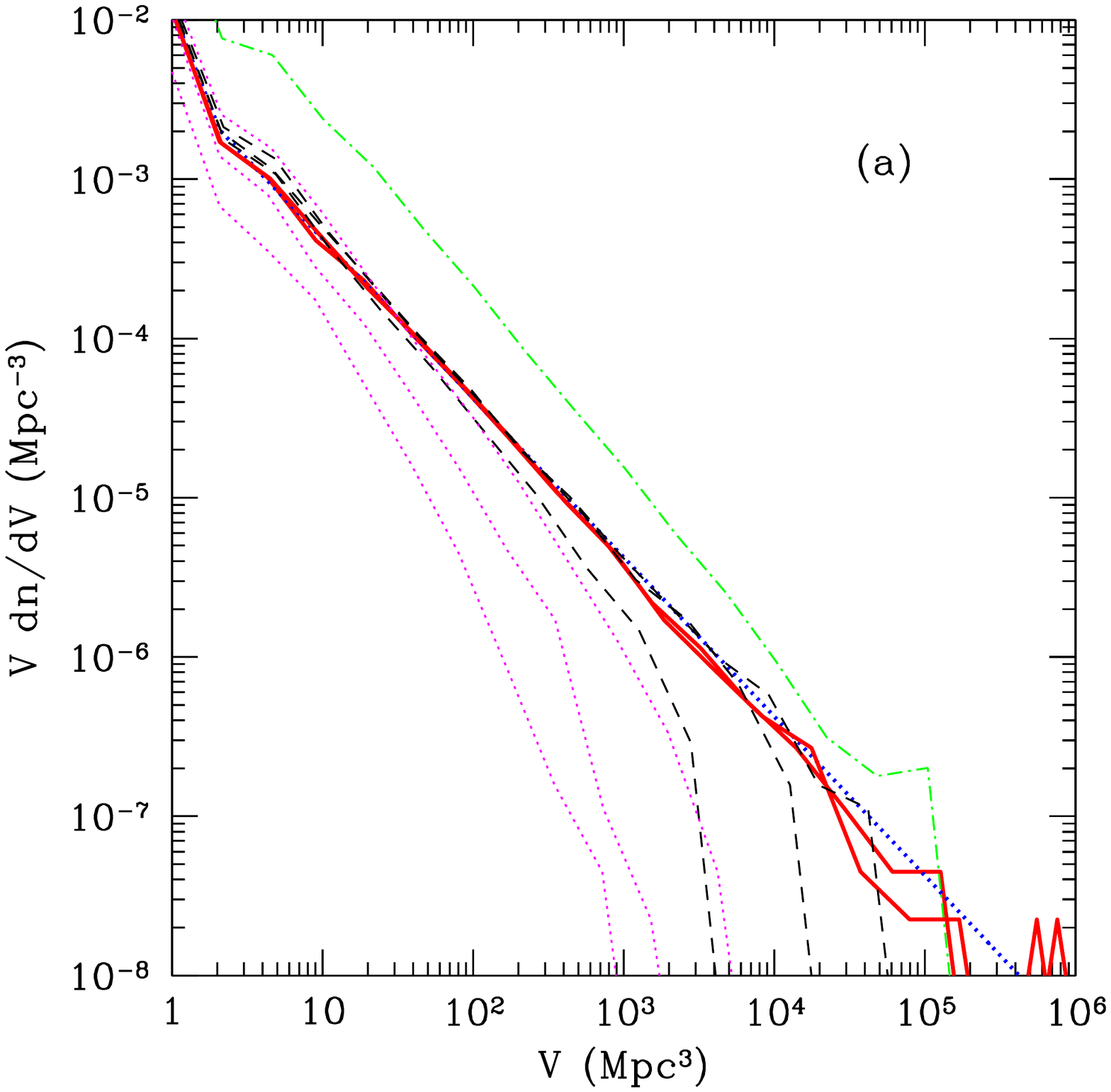}\includegraphics[width=\columnwidth]{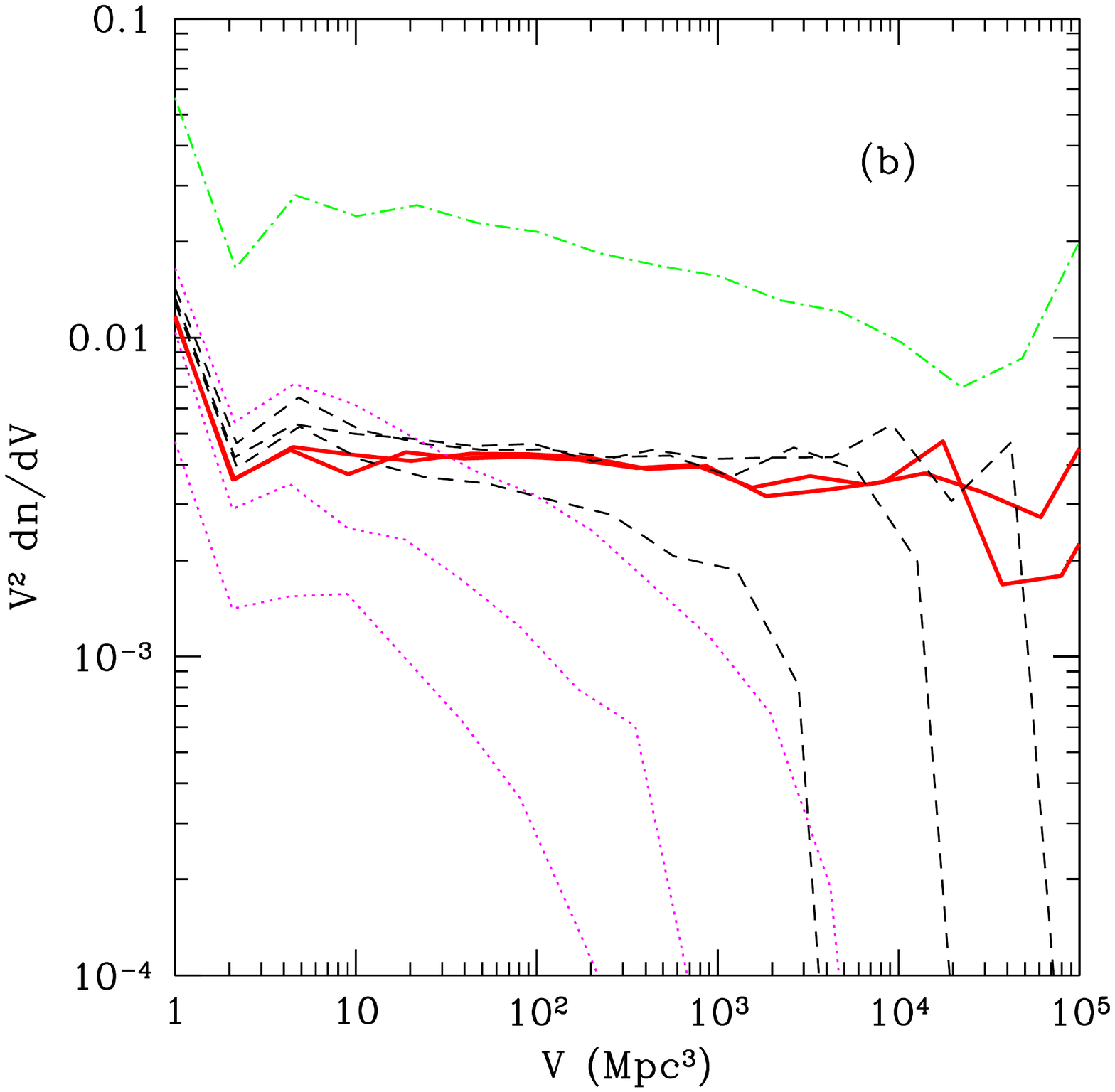}
\vskip -0.9in
    \caption{\emph{(a)} Size distribution of ionized regions in a $400^3$ box spanning $(400~\Mpc)^3$.  The red solid lines show the distribution just above and just below the percolation threshold, $x_{i,c} \approx 0.10$.  The blue dotted line shows a power law, $dn/dV \propto V^{-2}$,.  The dashed curves show the distributions below the threshold, while the dotted curves show distributions above the threshold (excluding the percolating cluster in the latter case). As in Fig~\ref{fig:random_size_distbn}, these follow roughly the same power law as the distribution at threshold before falling off at large sizes.  The green dot-dashed curve shows the simple cubic lattice result at its percolation threshold; note that it is slightly steeper.  \emph{(b)} Same, but weighted by bubble volume.}
    \label{fig:size_distbn}
\end{figure*}

\subsection{The Shapes of Ionized Regions} \label{shape}

Following \citet{shandarin06}, we can attempt to quantify the shapes of ionized regions by fitting ellipsoids to them. To do so, we demand that the fitted ellipsoid has an identical moment-of-inertia tensor to the ionized distribution, which uniquely specifies the three axes ($a$, $b$, and $c$) of the ellipsoid.  This means we weight cells by $r^2$, where $r$ is their distance from the center of mass of the particular ionized region of interest, when performing the fit; this guarantees that the centroid and principal axes of the ellipsoid match those of the bubble. The $r^2$ weighting is also motivated by another result from percolation theory: the average of $\VEV{r^2}$ over all cluster sites and all clusters is proportional to the square of the correlation length of the system, which diverges near the percolation threshold in a universal fashion. However, note that our definition does \emph{not} produce axes that correspond to the outer boundary of the network of tunnels and blobs that make up each cluster.  We find that the typical axis ratios of finite ionized regions are $b/a \approx 0.6$ and $c/a \approx 0.4$, for a wide range of sizes and ionized fractions.  In contrast, examination of Figure~\ref{fig:rendering} shows that the percolating cluster is not ellipsoidal, once it begins to fill the box: in that case we are no longer able to determine its shape in any quantitative fashion because of the periodic boundary conditions, but it appears to be a meaningless exercise anyway, because it is ``near" every point in the box.

Another useful measure, also introduced by \citet{shandarin06}, is the \emph{inverse porosity}.  We let $V_{\rm ellipsoid}$ be the volume of the best-fit ellipsoid; then the inverse porosity is the ratio of the true volume of the ionized region to $V_{\rm ellipsoid}$ and represents how efficiently the ionized bubble fills a ``regular" ellipsoidal volume.  We show the inverse porosity as a function of bubble size in Figure~\ref{fig:porosity}, excluding the infinite cluster.  Here the solid squares show the porosity very close to the percolation threshold ($x_i \approx 0.1$), while the triangles and hexagons show results somewhat farther from that value ($x_i \approx 0.06$ and $x_i \approx 0.14$, respectively).  The pentagons show the inverse porosity well above threshold ($x_i \approx 0.35$).  Note that a single pixel in this realization (which has $400^3$ cells across $[400 \Mpc]^3$) has a volume of $\sim 1 \Mpc^3$, so the discreteness of the lattice significantly affects our results for $V \la 100 \Mpc^3$; we have made no attempt to correct for this here.

For large (but finite) bubbles, the inverse porosity is significantly smaller than unity, indicating that the bubbles are much more complex than ellipsoids -- even filamentary regions could be reasonably well fit by stretched ellipsoids.  Instead, the finite regions must have many branches and sub-branches, so that they subtend large spaces without filling them. The shapes get more complex as the volume increases and more of these branches are added, so that the inverse porosity steadily decreases.  Note, however, that within each bin there is a large scatter in the shapes: the ``error bars" show the $2\sigma$ dispersion for selected volume bins. This dispersion is much larger than the trend in the average value, which also illustrates the complex shapes that reionization generates. We note that the shape distributions are not gaussian, with non-negligible skewness toward smaller values of the porosity and positive excess kurtosis.

However, comparison to the pentagons below the other points, which show the inverse porosity values derived from our random cubic lattice simulations, shows that the ionized regions are \emph{less} complex than randomly generated bubbles, at least for moderately-sized objects. This is likely a result of the filamentary structure and large-scale correlations that seed the sources.

One must bear in mind, however, that these finite ionized regions never contain more than a small fraction of the Universe: while they dominate the size distribution before and just after percolation, they contain only $x_i \la 0.1$ of the volume, and above the percolation threshold the vast majority of the ionized volume is in the infinite bubble.  The inverse porosity of the percolating cluster itself varies strongly with ionized fraction.  The rightmost square point in Figure~\ref{fig:porosity} shows the percolating cluster near threshold, with $V/V_{\rm ellipsoid} \approx 0.05$.  The Figure does not show the values for larger ionized fractions (as the bubbles are too large), but we find that the inverse porosity is roughly \emph{equal} to $x_i$ up to $x_i \sim 0.5$ and then increases rapidly to $V/V_{\rm ellipsoid} \approx 0.75$.  The first part is likely a reflection of the fact that the percolating cluster extends throughout the volume, but with many holes that are gradually filled in as the ionized fraction increases.

\section{Percolation and Ionized Bubble Sizes} \label{size-distbn}

The second salient feature of percolation processes is the ``universal" size distribution illustrated in Figure~\ref{fig:random_size_distbn}.  Figure~\ref{fig:size_distbn}\emph{a} shows size distributions from our $400^3$ box spanning $(400~\Mpc)^3$ at a variety of ionized fractions.  The solid lines show the distribution just above and just below the percolation threshold.  As expected, it is an almost pure power law: the dotted curve in between the cases shows $dn/dV \propto V^{-2}$, which provides a good fit over a wide range of scales.  This is slightly less steep than the corresponding distribution at the critical threshold in random percolation, shown by the dot-dashed curve here (and which has $dn/dV \propto V^{-2.18}$).  

Figure~\ref{fig:size_distbn}\emph{b} presents the same information in a different way, weighting the distribution by bubble volume.  (We truncate the $V$ axis at $10^5 \Mpc^3$ here because the curves become very noisy at larger volumes, where there is at most one ionized region per bin.) It thus shows the contribution to the \emph{total} ionized volume from each bubble scale.  Because $\tau \approx 2$ for the 21cmFAST simulations, at the percolation threshold all scales contribute equally: not only is there no ``characteristic" scale, but both large and small regions contribute equally to the total ionized structure.  The random case, on the other hand, has a slight decline with volume, indicating that small clusters are slightly preferred over large ones.

The other curves in the Figure show the distributions above and below $x_{i,c}$.  The dotted curves take $x_i \approx 0.14,\,0.35$, and $0.67$, respectively, while the dashed curves take $x_i \approx 0.072,\,0.0599$, and $0.043$, respectively.  Recall the general expectation that, at least below the percolation threshold, the distribution has $dn/dV \propto V^{-\tau}$, up to some cutoff size that becomes smaller as $|x_i - x_{i,c}|$ increases.  Below $x_{i,c}$, the distribution very closely follows the expected shape, with a steep cutoff appearing only at very large scales. Just as at threshold, bubbles on a wide range of scales contribute almost equally to the ionization pattern, and there is not really a preferred scale to the distribution, aside from the cutoff scale.  

However, above the threshold the distribution appears to differ, though it is difficult to determine just how much, as the range over which a power law applies is quite narrow and the transition toward the cutoff scale is less obvious.  It is clear that the decline above the cutoff is  gentler. We remind the reader that only a small fraction of the ionized gas is actually in these finite bubbles when $x_i > x_{i,c}$, and we see from the Figure that these bubbles also have rather small sizes.

\section{Reionization, Criticality, and Scaling Relations} \label{scaling}

Much of the literature on percolation theory concerns itself with the details near the phase transition, where the infinite cluster first appears and grows very rapidly.  It is near this point that the ``universal" properties of the transition dominate the behavior.  We emphasized two of these in \S \ref{cubic-lattice}: the growth of the percolating cluster itself, with $P_{\rm inf} \propto |x_i - x_{i,c}|^\beta$ near the percolation threshold, and the shape of the size distribution, $dn/dV \propto V^{-\tau}$ at threshold.  For random percolation in three dimensions, $\beta=0.41$ and $\tau=2.18$.  

When we introduce correlations to the process, however, these exponents change.  In particular, if the correlation function of the ionized fraction, $\xi_{xx}(r) \equiv [\VEV{x_i({\bf r}_i) x_j({\bf r}_j)} - x_i^2] \propto r^{-2H}$, then correlations change the universality class whenever  (in three dimensions) $H < 1.136$ \citep{weinrib83,weinrib84}.\footnote{In the percolation theory literature, the correlation function is usually written $g(r)$.}  Here, $H$ is the Hurst exponent.  In that case, the critical exponents are determined by the value of $H$.  While the correlation function is not a pure power law during reionization, it is reasonably close to such a form.  \citet{lidz08-constraint} have shown that, on moderately large scales, the fractional perturbation $\Delta^2 \propto k^3 P(k) \propto k^{n}$, with $n \sim 1$--2 (0.5) during the early (late) stages of reionization. For a pure power law, this would correspond to $\xi_{xx} \propto r^{-n}$ with $H \sim 0.25$--$1$, well within the regime in which correlations affect the structure throughout reionization.  Nevertheless, even in this regime there are algebraic relations between the critical exponents that appear to be universal, though they do depend upon the dimensionality of the system (e.g., \citealt{saberi15}).  The numerical values of the exponents have only been estimated in a few specific cases, and there is no general rule for determining them.  (Even in the uncorrelated problem, only a very few cases permit rigorous calculations of the exponents. Most configurations require extensive numerical experiments.)

The growth of $P_{\rm inf}$ is shown in Figure~\ref{fig:order_param}.  While both the reionization and random cases show a rapid increase in $P_{\rm inf}$ at the percolation threshold, it is clear that the rate of increase slows down more quickly for the reionization case, leading to a much longer tail of ionized regions \emph{outside} the infinite cluster for the 21cmFAST simulations. Intuitively, this is because of clustering: while most of the ionized regions easily connect along filaments, the isolated H~II regions that do form within voids have a harder time connecting to that established network. Mathematically, this tail manifests itself as $\beta \approx 0.8$, about twice the value for a random process.  Interestingly, our results are also consistent with the value of $\beta=0.76$ found by \citet{shandarin10} for percolation of the density field in a cosmological simulation of the $\Lambda$CDM model at $z=0$.  They found, however, that the initial gaussian random field in that model has $\beta \sim 0.5$--$0.6$.  

We have also seen that the power-law slope $\tau$ of the size distribution is shallower than that for the random lattice, at least at and below the percolation threshold.  Again, this is a natural result of clustering, which will help larger ionized bubbles to grow more rapidly than in the random case.  Moreover, our value is in good agreement with the measurement of \citet{shandarin06} from simulations of large-scale structure.  They found $\tau \approx 1.95$ for voids in a gaussian random field, and for the power law portion of voids in a nonlinearly evolved universe, which is also consistent with our distribution.  This suggests that it is the correlations intrinsic to the density field (and hence sources) that set this slope, rather than any particular details of the reionization process.

%FIGURE: Redshift dependence
\begin{figure}
	\includegraphics[width=\columnwidth]{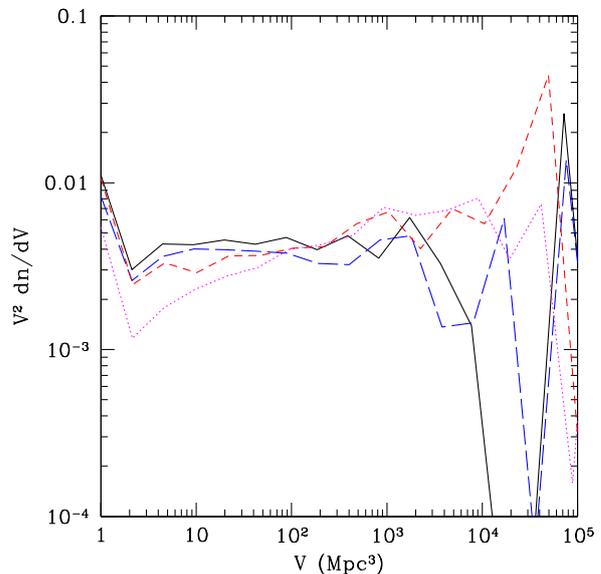}
\vskip -0.9in
    \caption{Volume-weighted size distributions near $x_{i,c}$ at $z=7,\, 10,\, 15,$ and 20 (long-dashed, solid, short-dashed, and dotted curves, respectively).  All are evaluated in $200^3$ boxes spanning $(200~\Mpc)^3$. As in Fig.~\ref{fig:size_distbn}, we exclude the percolating cluster.}
    \label{fig:z_dependence}
\end{figure}

\subsection{Redshift Dependence} \label{size-z}

So far we have focused on ionization fields at $z=10$.  We now briefly examine the behavior at different redshifts: \citet{shandarin06} found that, the percolation properties of the cosmological density field evolved somewhat once the relevant scales became nonlinear.  Our semi-numeric simulations do not follow the IGM far into the nonlinear regime, but it is still represented by the source distribution (most of which has gone nonlinear). 

Numerically, we find that the percolation thresholds themselves change somewhat over the relevant redshift interval: at $z=15$ and $z=20$ they are $x_{i,c} \approx 0.12$, while at $z=7$ it is $x_{i,c} \approx 0.08$.  This spread is comparable to the scatter from cosmic variance in Figure~\ref{fig:order_param_cosvar}. However, here we have made our measurements from a single realization of the density field at all redshifts, so the spread is nevertheless significant.  We hypothesize that the difference is due to evolution in the source number density. At a fixed neutral fraction, the $z=20$ box relies on larger ionized regions around fewer sources. Connections between these regions are more subject to random fluctuations than at lower redshifts, when there are many more sources strung out along the filaments.

In Figure~\ref{fig:z_dependence}, we show the volume-weighted size distributions near the percolation thresholds at a range of redshifts: $z=7,\, 10,\, 15,$ and 20 for the long-dashed, solid, short-dashed, and dotted curves, respectively.  There is not a tremendous difference amongst the redshifts, but the $z=15$ and especially the $z=20$ curves are slightly steeper than the lower redshift points.  This supports our hypothesis of stronger clustering affecting the process, because that should also cause bigger ionized regions to appear.  

\section{Discussion} \label{discussion}

In this section, we describe some of the implications of percolation theory for studying and observing cosmic reionization.

\subsection{The phases of reionization} \label{phases}

For well over a decade, astronomers have conceptually divided reionization into three stages (see, e.g., \citealt{gnedin00, loeb13}): ``pre-overlap," in which ionized regions grow in isolation; ``overlap," in which the mergers of ionized bubbles dominate the dynamics of the field; and ``post-overlap," in which the last remnants of neutral material is ionized.  Percolation theory offers a more rigorous definition of these eras in terms of infinitely large ionized and neutral regions while preserving their intuitive meaning.

To this point, we have treated the appearance of an infinitely large ionized region as a percolation process.  But one can equally well treat the \emph{disappearance} of the neutral gas in the same way.  In random percolation, the time-reversed process would be identical; here, however, the evolving large-scale correlations of the initial and final phases of reionization will introduce some differences.

%FIGURE: Order parameter for neutral regions
\begin{figure}
	\includegraphics[width=\columnwidth]{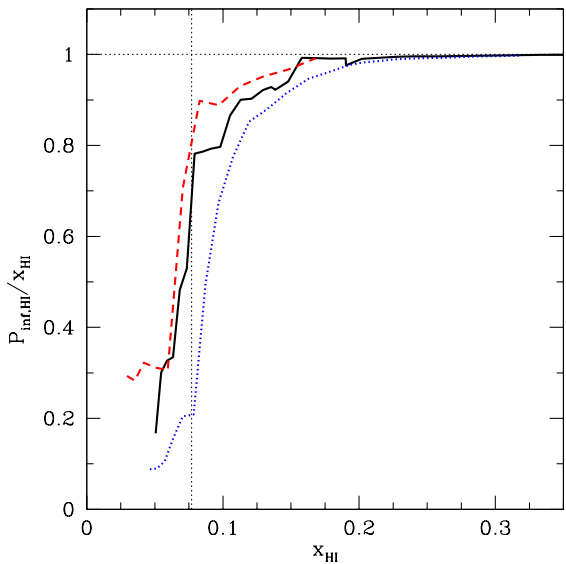}
\vskip -0.9in
    \caption{Order parameter $P_{\rm inf,HI}$ for ``time-reversed" percolation of neutral gas, scaled to the occupation fraction $x_{\rm HI}$, of that phase.  Above the percolation threshold, the curves show the fraction of the neutral volume inside the spanning cluster, while below the threshold they show the fraction inside the largest cluster. The solid and dashed curves show results for two different realizations of a 400$^3$ box spanning $(400 \Mpc)^3$, with ionizing photons allowed to reach $R_{\rm max}=30$~Mpc. The dotted curve shows the same realization as the dashed curve, but it takes $R_{\rm max}=10$~Mpc. The dotted vertical curve show the percolation threshold for the $R_{\rm max} = 30 \Mpc$ curves.}
    \label{fig:order_param_neutral}
\end{figure}

Our present understanding of the end of reionization is fairly crude, as it depends on detailed interactions between the ionizing sources and small-scale absorbers in the IGM.  The implementation in 21cmFAST is even cruder, simply imposing a maximum size over which any given source can influence the ionization field.  We therefore will not focus on the detailed properties of this percolation process here, but in Figure~\ref{fig:order_param_neutral} we show the order parameter $P_{\rm inf,HI}$ for this second percolation transition, scaled this time to the total neutral fraction $x_{\rm HI}$.  The solid and dashed curves compare two different realizations of the process, both in 400$^3$ boxes spanning $(400 \Mpc)^3$.  In these cases, we use 21cmFAST's default choice of $R_{\rm max} = 30 \Mpc$.  A percolation transition clearly occurs, with the fraction of neutral gas in the infinite cluster approaching unity even faster than the corresponding ionized gas transition.  However, because $R_{\rm max}$ is is large (and imposed discontinuously), the curves are much less smooth than their analogs.  In both cases, $x_{{\rm HI},c} \approx 0.077$.  This is smaller than $x_{i,c}$ for ionization, but given the different correlations in the two regimes the difference is not surprising.

The dotted curve shows the progression if $R_{\rm max}=10 \Mpc$, within the same realization of the density field as in the dashed curve.  This is much smoother -- as there are many more quasi-independent volumes in the simulation box -- but percolation occurs at nearly the same point as the other curves.  The primary difference is that the largest cluster below the percolation threshold is significantly larger for $R_{\rm max}=30 \Mpc$, probably because that case has smaller ``dynamic range" (c.f. the effects of box size on large regions below the percolation threshold in Figure~\ref{fig:order_param}).

We will refrain from analyzing this second transition in detail until a more satisfactory physical model of absorption is available \citep{choudhury09,sobacchi14,davies15-uvb}.  But we have demonstrated the existence of \emph{two} percolation transitions, which divide the history of reionization into three phases:

\emph{(i)} During the \emph{pre-overlap phase}, ionized gas is confined to finite discrete regions.  The distribution of these regions is closely related to that of the sources.  This phase persists until $x_i \approx 0.1$. Importantly, even in this phase the ionized regions can grow very large, and there is no well-defined characteristic scale when they are allowed to take arbitrary connected shapes.

\emph{(ii)} The \emph{overlap phase} commences with the percolation of a unique infinitely large cluster, which then undergoes a rapid series of mergers with existing large ionized regions.  Subsequently, this infinite region grows nearly linearly with the ionized fraction and includes nearly all the ionized gas, with the discrete remnants decreasing in characteristic size as reionization progresses.  At the same time, nearly all of the neutral gas is part of its own unique infinitely-large structure, which intertwines with the percolating ionized region in complex ways.

\emph{(iii)} At $x_i \ga 0.9$, the infinite neutral region fragments and reionization enters the \emph{post-overlap} phase.  The percolating neutral cluster breaks up into pieces, likely still very large, which are gradually consumed.  This phase is controlled by IGM absorption, which affects the range over which sources can reach the neutral gas and so determines the size distribution of the neutral regions.

The merger tree approach to H~II regions developed by \citet{chardin12} complements this picture, demonstrating that the number of discrete H~II regions peaks at about the time the overlap phase begins, and that overlap is driven primarily by the merger of discrete ionized bubbles with the percolating region.

\subsection{Percolation and observations} \label{obs}

While this paper is primarily concerned with the fundamental theory of the reionization process, it has some clear implications for observations, and in particular imaging of the ionization field through the 21-cm line. The most significant result is the simultaneous existence of a unique percolating cluster in the ionized medium and a unique percolating cluster in the neutral medium throughout the bulk of reionization ($x_i \sim 0.1$--$0.9$), both of which contain the vast majority of the volume of their respective phases. This alters the prevailing picture of the structure of reionization, which is usually described as measuring the ``size distribution" of discrete ionized regions, which we have shown to contain only a very small fraction of the ionized and neutral volume. While there are certainly characteristic scales in the ionization field -- as shown by, e.g., \citet{mesinger07} and examined in detail by \citet{lin15} -- these do not mark distinct ionized bubbles but are instead more subtle features of the distribution.  Regardless, they contain vital information about the physics of reionization, including both the sources and sinks of ionizing photons, and they should therefore be a prime target both for observations and more sophisticated theoretical understanding. The stochastic percolation process undoubtedly affects these scales as well, and the complexity of the resulting structures also implies that measured characteristic sizes must be carefully defined, as they will depend on the measurement process itself \citep{lin15}.

The model-independent nature of the percolation transitions also implies that they can serve as useful signposts of the reionization process. An infinite ionized region \emph{must} exist after $x_i \approx 0.1$, and its appearance (provided that it can be detected) would be a robust test of the observations.  It remains to be seen, however, how well such a region can be identified in noisy, low-resolution data, as it is far from a simple spherical distribution and  smoothing will not necessarily help to identify it when it is so deeply intertwined with the neutral gas.  Still, it may be possible to identify large neutral regions in noisy data through statistical analysis of connected regions. The spatial resolution required to identify the percolating region will depend sensitively on how close to $x_{i,c}$ one looks (and will be the subject of future work). Nevertheless, Fig.~\ref{fig:zoom_structure} suggests that the ``skeleton" of the cluster can be detected with a resolution of several comoving Mpc, even fairly close to the threshold. On the other hand, the dangling ends that fill most of the volume have more tenuous connections and will require better resolution to reliably connect to the percolating region.

The power-law size distribution of the discrete ionized (and presumably neutral) regions implies that very large features will be visible even in the early and late stages of the process.  In particular, Figure~\ref{fig:size_distbn} shows that bubbles grow to very large scales even when $x_i < 0.1$. Although we have not shown it explicitly (because 21cmFAST does not include enough physics to accurately model the very late stages), a similar effect seems to hold for the neutral regions.  These very large neutral regions are likely the easiest features to identify individually of any throughout reionization, so our work provides reason for optimism in regards to 21-cm imaging.

Additionally, the complex, three-dimensional structure of the ionized regions demands that we treat imaging data three-dimensionally in order to understand the topology of reionization: otherwise, as in Figure~\ref{fig:slices}, the connectedness of the underlying distribution will be completely obscured.  It will require more sophisticated imaging techniques than generally utilized in astronomy, where three-dimensional information is so often lacking (and often not even relevant).

Finally, the existence of the percolating cluster and large ionized bubbles have indirect implications for other observations. For example, the Lyman-$\alpha$ lines of galaxies will be extinguished if the galaxy is too close to neutral gas (e.g., \citealt{haiman02-lya,santos04,furl06-lya}). Previous work has shown some surprising trends in this absorption, especially the fact that the absorption is mostly independent of the galaxy mass, despite the increased clustering of massive galaxies \citep{mcquinn07-lya, mesinger08-lya, iliev08-lya}. This may be related to the fractal structure of the percolating cluster.

\subsection{Percolation and theoretical models} \label{theory}

The first discussions of the structure of the ionization field treated it as composed of discrete bubbles \citep{shapiro87, furl04-bub}, and that viewpoint has informed much of our discussion of the models, even though evidence for very large percolating structures has been found \citep{iliev06-sim, chardin12}.  Here we have shown that -- while the physical arguments at the foundation of those models still appear reasonable -- discrete bubbles (or neutral regions) are only relevant at the very beginning (or end) of reionization.

The problem with these analytic models is simply the assumption of spherical symmetry, which is necessary to make the model tractable but is far too limiting when describing the complex structures of the ionization field (even in the absence of an inhomogeneous IGM).  Two-dimensional slices of the ionization structure (such as Fig.~\ref{fig:slices}) disguised this weakness and led our collective intuition astray.

Given the difficulties of modeling radiative transfer and ionization structures using analytic models, most work on reionization uses numerical or semi-numeric simulations, both of which have increased vastly in sophistication over the past fifteen years.  We have used the simplest version of these methods in this paper and demonstrated that percolation has always been present: our results do not require a substantial change in the modeling techniques but a more sophisticated analysis of their results.  

As an example, the dominance of the percolating cluster may explain the early importance of inhomogeneous recombinations to reionization \citep{choudhury09, sobacchi14}.  As originally argued by \citet{furl05-rec}, IGM absorption should only become significant once the mean free path becomes shorter than the typical distance from a source to the neutral gas.  If the ionized regions remained discrete, recombinations would therefore not become important until late in reionization.  However, the fact that the majority of the ionized volume is inside a single region as early as $x_i \sim 0.1$ (albeit one with a complex structure) provides a natural explanation for this effect being important much earlier in the process.

Moreover, the quasi-random generation of large ionized regions from the overlap of ``distinct" source regions -- the analog to the growth of large clusters on our random cubic lattice example -- may also explain why the ionized bubbles in simulations are larger than those predicted by the analytic excursion set model \citep{iliev06-sim, friedrich11}. \citet{lin15} have demonstrated that, when bubbles are detected with different criteria (such as the ``watershed algorithm," or by measuring the distribution of path lengths within ionized regions), a clear characteristic scale does emerge -- albeit one several times larger than that predicted by a naive application of the excursion set theory (see, e.g., \citealt{paranjape14} for modifications to excursion set theory which potentially improve agreement). The origin of this characteristic scale is not yet clearly understood, but it likely results from a combination of the ``natural" scale imprinted by source clustering, stochastic overlap from percolation, and (once properly modeled) the propagation of ionizing photons through the IGM. Understanding the interaction of these scales with each other will be essential to extracting physical information from maps of the neutral and ionized gas.

Our results also offer some cautionary notes for attempts to simulate the reionization process. Percolation is an infamously difficult problem to treat accurately with finite simulations, especially near the percolation transition, as the size distribution becomes a pure power law and hence scale free. Because the percolation threshold occurs very early in the process of reionization ($x_i \approx 0.1$), even simulations of the early phases of reionization must have a high enough dynamic range to encompass a large range of scales, if they wish to properly characterize the structure of the ionization field.  It is an \emph{intrinsically} large-scale process, and treating ionized regions in isolation is never a good approximation.  A similarly large range of scales is essential for the end stages of reionization, especially because the rare large neutral regions will be the easiest to observe. For calculations with other foci, the requirements are likely more modest, as even in the presence of the percolating cluster a well-defined characteristic scale for the mean free path of photons does exist \citep{lin15}. However, one must certainly take care to ensure that both the mean and variance of the radiation field have converged as a function of box size once percolation occurs.

\subsection{Reionization and large-scale structure} \label{lss}

As emphasized earlier, percolation theory has been applied to understanding the large-scale structure of the Universe, especially with regard to the voids and filamentary structure that fills the bulk of the volume \citep{shandarin06, shandarin10}. There are some key differences between the cases, however. First, the percolation parameter in large-scale structures is a density threshold, which is varied at a fixed redshift to study how the lower-density material fills space. This is conceptually convenient but difficult to observe and interpret given its continuum nature. In our case, the percolation aspect is much more obvious, as the division between neutral and ionized gas is physically relevant and directly observable.  Second, previous models have shown that the reionization process is primarily driven by \emph{linear} perturbations which, at these early redshifts, are accurate on the large scales relevant to the problem (though recombinations affect this expectation on some level). Large-scale structure in the nearby Universe, however, has crucial nonlinear effects which complicate the mapping. Third, reionization is more obviously non-local, in that distant galaxies affect the state of any given cell.

Nevertheless, we have found that our results largely mirror those of \citet{shandarin10}.  In particular, both cases have a percolation threshold at filling factors near $\approx 0.1$ and both cases have similar power laws for the size distribution of discrete regions and the growth of the percolating cluster. Thus the ``trappings" of reionization are of secondary importance, at least within the limits of our simulations: the percolation behavior is driven not by the detailed properties of the ionizing sources but by the underlying cosmology. It remains to be seen whether this is true in more sophisticated models, especially when inhomogeneous recombinations affect the large-scale behavior.  

\subsection{Percolation and the Euler characteristic} \label{euler}

To date, there has been relatively little discussion of the structure of ionized and neutral regions during reionization.  The most popular measure -- because it has been applied numerous times to large-scale structure -- has been the Euler characteristic or, equivalently, the genus number \citep{gleser06, lee08, friedrich11, hong14, wang15}.  The genus number $g$ is the number of complete cuts one can make through an object without dividing it into disconnected parts, and it is related to the Euler characteristic $V_3$ (also known as the third Minkowski functional) through $g = 1 - V_3$.  The Euler characteristic is the integral of the Gaussian curvature over the surface and is proportional to
\begin{equation}
\mbox{(no. of parts)} \, - \, (\mbox{no. of tunnels}) 
\, + \, (\mbox{no. of cavities}).
\label{eq:euler}
\end{equation}
Generically, $V_3(x_i)$ varies significantly throughout reionization (see, e.g., Fig.~8 of \citealt{friedrich11}, or the two-dimensional version in Fig.~4 of \citealt{wang15}). 

Given that the Euler characteristic depends on the number of independent ionized regions, it is easy to see the origin of this behavior: during the long overlap phase, there is really only one enormous ionized region, with a complex structure interlaced with a single enormous neutral region. Both discrete ionized regions and discrete cavities are rare, as nearly all of the volume is inside these infinitely large regions.  Only toward the end of the process, past the percolation threshold for the neutral gas, does the number of discrete cavities increase, pushing the Euler characteristic to positive values.

The Euler characteristic and genus number therefore serve as useful global measures of the percolation topology, and if they can be measured in real data (which has not yet been examined closely, but appears difficult) can offer somewhat model-independent measures of the reionization process.

\section{Conclusions} \label{conc}

In this paper, we have reviewed reionization as a percolation process.  We have shown that it exhibits a pair of percolation phase transitions. In the first (at $x_i \sim 0.1$), the bulk of the ionized gas is quickly incorporated into a unique infinitely large ionized region, while during the second (at $x_i \sim 0.9$) a unique neutral region breaks into discrete neutral regions.  We have also shown that the rate at which this infinite region grows follows a power law near the percolation threshold, as expected for a phase transition exhibiting universality.  However, the exponent in this power law differs from that in uncorrelated three-dimensional percolation, thanks to large-scale correlations in the sources driving reionization. Instead, the critical behavior is (unsurprisingly) similar to that found in studies of large-scale structure in the density field \citep{shandarin10}.

Meanwhile, below the percolation threshold the discrete ionized regions (when defined as simply connected volumes, regardless of their shapes) also follow a power law size distribution, with nearly equal filling factors per logarithmic interval in bubble volume up to some characteristic size (which increases as the Universe approaches the critical point). Analytic models, which make strong assumptions about symmetry, have in the past predicted a well-defined characteristic size for ionized regions (e..g., \citealt{furl04-bub}). In reality, the stochastic nature of percolation forces the true size distribution -- when all connections are included - to approach a scale-free form at the threshold. Still, it appears that \emph{a} characteristic scale appears in the size distribution, when constructed on a more practical level, thanks to the way in which regions connect to the infinite cluster through narrow channels \citep{lin15}. 

Understanding the true shapes of bubbles, and how different processing techniques ``filter" these shapes, promises to be important for successfully interpreting images, and possibly statistical observations, of the ionized topology. For example, the ionized regions, in both 3D renderings and in 2D slices, do appear to have a characteristic size, which represents some combination of the composite ionizing efficiency of a typical overdensity of galaxies, the stochastic overlap of such regions, and the inhomogeneous absorption of the IGM.  It is crucial to understand the physical origin of this scale and its connection to the percolation process. One could, very crudely, think of each such characteristic scale-sized region as a cell in a larger-scale percolation problem, with IGM absorption reducing the effect of correlations beyond the typical mean free path. 

The percolation transitions provide precise meaning for the intuitive description of reionization that has existed for many years \citep{gnedin00}, the pre-overlap, overlap, and post-overlap phases. The pre-overlap phase simply describes the Universe before the appearance of the infinite ionized region, while post-overlap describes the Universe after the infinite neutral region fragments into discrete pieces. The overlap phase -- which in reality constitutes the \emph{bulk} of reionization, encompassing $0.1 \la x_i \la 0.9$ -- sees the ionized and neutral gas almost entirely contained in just two distinct, delicately intertwined regions. 

Here we have shown that treating reionization as a percolation process transforms our understanding of the topology and morphology of ionized and neutral gas. For clarity, we have focused on a specific set of reionization calculations. In the future, a more thorough exploration of parameter and algorithmic dependencies will allow us to understand the sensitivity of the percolation process to our modeling. This will determine whether the percolation transition provides a robust signpost to confirm our analysis of e.g., 21-cm images, or alternatively whether it provides a new probe of the underlying physics of high-redshift sources.

\section*{Acknowledgments}

SRF thanks R. Mebane for discussions during the early stages of this project, and we thank Y. Lin for helpful discussions throughout. SRF also thanks an NSF review panel for the impetus to complete this project. This research was completed as part of the University of California Cosmic Dawn Initiative. We acknowledge support from the University of California Office of the President Multicampus Research Programs and Initiatives through award MR-15-328388. SRF was also partially supported by a Simons Fellowship in Theoretical Physics and thanks the Observatories of the Carnegie Institute of Washington for hospitality while much of this work was completed. We thank T. Fricke for making a C implementation of the Hoshen-Kopelman algorithm publicly available, A. Fog for his public implementation of the Mersenne Twister algorithm, and the VisIt team for making their three-dimensional rendering code publicly available. 

%%%%%%%%%%%%%%%%%%%%%%%%%%%%%%%%%%%%%%%%%%%%%%%%%%



%\bibliographystyle{mnras}
%\bibliography{Ref_composite} % if your bibtex file is called example.bib

%%%%%%%%%%%%%%%%%%%%%%%%%%%%%%%%%%%%%%%%%%%%%%%%%%

%%%%%%%%%%%%%%%%% APPENDICES %%%%%%%%%%%%%%%%%%%%%

%\appendix


%%%%%%%%%%%%%%%%%%%%%%%%%%%%%%%%%%%%%%%%%%%%%%%%%%

% Don't change these lines
\bsp	% typesetting comment
\label{lastpage}
\end{document}